# Propionamide (C$_2$H$_5$CONH$_2$): The largest peptide-like molecule in space

Juan Li,[1, 2] Junzhi Wang,[1, 2] Xing Lu,[3] Vadim Ilyushin,[4, 5] Roman A. Motiyenko,[6] Qian Gou,[7] Eugene A. Alekseev,[4, 5] Donghui Quan,[8, 9] Laurent Margule`s,[6] Feng Gao,[10, 11] Frank J. Lovas,[12, 13] Yajun Wu,[1, 2] Edwin Bergin,[14] Shanghuo Li,[15] Zhiqiang Shen,[1, 2] Fujun Du,[16, 17] Meng Li,[7] Siqi Zheng,[1, 2, 18] and Xingwu Zheng[19]

[1]*Department of Radio Science and Technology, Shanghai Astronomical observatory, 80 Nandan RD, Shanghai 200030, PR China*

[2]*Key Laboratory of Radio Astronomy, Chinese Academy of Sciences, PR China*

[3]*National Astronomical Observatory of Japan, 2-21-1 Osawa, Mitaka, Tokyo, 181-8588, Japan*

[4]*Institute of Radio Astronomy of NASU, Mystetstv 4, 61002 Kharkiv, Ukraine*

[5]*Quantum Radiophysics Department, V. N. Karazin Kharkiv National University, Svobody Square 4, 61022 Kharkiv, Ukraine*

[6]*Laboratoire de Physique des Lasers, Atomes, et Molécules, UMR CNRS 8523, Université de Lille I, F-59655 Villeneuve d'Ascq Cédex, France*

[7]*School of Chemistry and Chemical Engineering, Chongqing University, Daxuecheng South Rd. 55, 401331, Chongqing, PR China*

[8]*Xinjiang Astronomical Observatory, Chinese Academy of Sciences, 150 Science 1-Street, Urumqi 830011, PR China*

[9]*Department of Chemistry, Eastern Kentucky University, Richmond, KY 40475, USA*

[10]*Hamburger Sternwarte, Universität Hamburg, Gojenbergsweg 112, 21029, Hamburg, Germany*

[11]*Max Planck Institute for Extraterrestrial Physics (MPE), Giessenbachstr. 1, 85748 Garching, Germany*

[12]*11615 Settlers Circle, Germantown MD 20876, USA*

[13]*Guest worker at Sensor Science Division, National Institute of Standards and Technology, Gaithersburg, MD 20899 USA*

[14]*Department of Astronomy, University of Michigan, Ann Arbor, MI 48109, USA*

[15]*Korea Astronomy and Space Science Institute, 776 Daedeokdae-ro, Yuseong-gu, Daejeon 34055, Republic of Korea*

[16]*Purple Mountain Observatory, Chinese Academy of Sciences, Nanjing 210034, PR China*

[17]*School of Astronomy and Space Science, University of Science and Technology of China, Hefei 230026, PR China*

[18]*University of Chinese Academy of Sciences, 19A Yuquanlu, Beijing 100049, PR China*

[19]*School of Astronomy and Space Science, Nanjing University, Nanjing 210093, PR China*

Corresponding author: Juan Li, Junzhi Wang

lijuan@shao.ac.cn, jzwang@shao.ac.cn

ABSTRACT

Peptide bonds, as the molecular bridges that connect amino acids, are crucial to the formation of proteins. Searches and studies of molecules with embedded peptide-like bonds are thus important for the understanding of protein formation in space. Here we report the first tentative detection of propionamide ($C_2H_5CONH_2$), the largest peptide-like molecule detected in space toward Sagittarius B2(N) at a position called N1E that is slightly offset from the continuum peak. New laboratory measurements of the propionamide spectrum were carried out in the 9-461 GHz range, which provide good opportunity to check directly for the transition frequencies of detected interstellar lines of propionamide. Our observing result indicates that propionamide emission comes from the warm, compact cores in Sagittarius B2, in which massive protostellars are forming. The column density of propionamide toward Sgr B2(N1E) was derived to be $1.5 \times 10^{16}$ cm$^{-2}$, which is three-fifths of that of acetamide, and one-nineteenth of that of formamide. This detection suggests that large peptide-like molecules can form and survive during star-forming process and may form more complex molecules in the interstellar medium (ISM). The detection of propionamide bodes well for the presence of polypeptides, as well as other complex prebiotic molecules in the ISM.

1. INTRODUCTION

Proteins, as the building blocks of living cells, are an essential component of living systems on Earth. Proteins are polymers of amino acids joined together by the peptide bond, -NHCO-. Due to the high molecular weight and extremely low gas-phase abundance, detection of proteins in the interstellar medium (ISM) at the current stage of development of observation facilities looks like a formidable task. Therefore, molecules with peptide-like bonds (Fig. 1) are of particular interest for our understanding of possible routes of protein formation in space (Halfen et al. 2011; Kaiser et al. 2013; Belloche et al. 2017).

The number of peptide-like molecules found so far in space is quite limited. The simplest molecule containing a peptide-like bond, formamide ($NH_2CHO$), was detected in Sagittarius B2 (Sgr B2) in the 1970s (Rubin et al. 1971), and later in various other sources, including the solar-type protobinary IRAS 16293-1622 (Coutens et al. 2016), the protostellar shock region L1157-B1 (Mendoza et al. 2014), comet 67P/Churyumov-Gerasimenko (Altwegg et al. 2017) and so on. About 30 yr later, its homologue (homologue means molecules that have the same functional group -NHCO-, but with additional methyl group, $CH_3$-), acetamide ($CH_3CONH_2$), was detected in Sgr B2(N) (Hollis et al. 2006; Halfen et al. 2011). N methylformamide ($CH_3NHCHO$) and urea ($NH_2C(O)NH_2$) were also identified in Sgr B2(N) (Belloche et al. 2017, 2019). Urea was also detected in a quiescent molecular cloud in the Sgr B2 complex (Jiménez-Serra et al. 2020). Recently, a number of the amide molecules, including $NH_2CHO$, $CH_3CONH_2$, and $CH_3NHCHO$, were detected in massive cores NGC 6334I (Ligterink et al. 2020). These observations suggest that peptide-like molecules might be widespread in space. Whether larger molecules with a peptide-like bond could form in the ISM is still an open question. In this context, the search for more complex interstellar peptide-like molecules is a key to understanding the growth of peptides.

Propionamide ($C_2H_5CONH_2$), a complex alkyl amide with twelve atoms (Fig. 1) (Marstokk et al. 1996), is the next member of the amide chemical family after formamide and acetamide. Various linear and cyclic aliphatic amides, such as monocarboxylic acid amides, dicarboxylic acid monoamides, hydroxy acid amides, carboxy lactams, lactims, and N-acetyl amino acids, have been found in the Murchison meteorite (Cooper & Cronin 1995). Specifically, propionamide was positively identified in the Murchison meteorite (Cooper & Cronin 1995), suggesting an interstellar origin of this molecule. Halfen et al. (2011) suggested an interstellar search for propionamide. They pointed out that the identification of this species in interstellar gas will be challenging, especially given the high density of rotational levels. However, continuous upgrades to observation facilities open new horizons for such searches.

Sgr B2, the massive star-forming region located close to the Galactic Center, has been found to be one of the best hunting grounds for complex organic molecules in the ISM (Li et al. 2017, 2020). Many prebiotic molecules have been detected in this source, such as the branched molecule i-$C_3H_7CN$ (Belloche et al. 2014) and the chiral molecule $CH_3CHCH_2O$ (McGuire et al. 2016). Sgr B2 contains two main sites of star formation, Sgr B2(N) and Sgr B2(M), both of which host several dense, compact, hot cores (Bonfand et al. 2017).

In this paper, we report the tentative detection of $C_2H_5CONH_2$ toward Sgr B2(N). The new laboratory spectroscopic study of propionamide spectrum is discussed in Section 2. The details of the observations and data reduction are described in Section 3. The observation results are presented in Section 4, along with the derivation of column densities of other related molecules of interest. The possible formation mechanisms of propionamide are discussed in Section 5. The results are summarized in Section 6.

2. SPECTROSCOPY

As a preparatory step of our ISM search, we performed a new laboratory spectroscopic study of the propionamide microwave spectrum. Previously available data ([Marstokk et al. 1996](#)) were limited in the 21.4– 39 GHz range, and extrapolations to the 84– 114 GHz range, where we performed our astronomical search, cannot be taken as reliable. The new laboratory measurements of the propionamide spectrum were carried out in the 9– 461 GHz range, providing a good opportunity to check directly for the transition frequencies of detected interstellar lines of propionamide. The rotational transitions belonging to the ground (v=0), first excited skeletal torsion (v=1), and first excited methyl torsion vibrational states were analyzed providing an accurate sets of constants for spectral predictions.

From a spectroscopic point of view, propionamide is a molecule with methyl torsional large amplitude motion (red arrow at Fig. 1e) hindered by a relatively high potential barrier ($\sim 760$ cm$^{-1}$). There are three factors that complicate analysis of its rotational spectrum. First, the barrier is rather high, and even in the first excited methyl torsion state $v_{29} = 1$, A – E splittings due to large amplitude methyl torsion motion are of the order of several MHz only. In the ground state, the A – E splittings are resolved mainly for high $K_a$ transitions. Thus, there is not much information on the methyl torsion motion obtained from the rotational spectrum of the molecule and the methyl torsional parameters in the Hamiltonian model are highly correlated. Second, there is a low lying skeletal torsion vibration mode $v_{30}$ in this molecule (blue arrow in Fig. 1e), which is very likely to interact both with the ground vibrational state and methyl torsion mode $v_{29}$. Finally, there are additional hyperfine splittings due to the non-zero electric quadrupole moment of the nitrogen atom, which by order of magnitude are comparable in many cases with methyl torsion A – E splittings in this molecule. The presence of two low lying vibrational modes (skeletal torsion $v_{30}$ with fundamental frequency of 45(7) cm$^{-1}$ (Marstokk et al. 1996), and methyl torsion $v_{29}$ with fundamental frequency of 179.6 cm$^{-1}$) and A-E methyl torsion plus quadrupole splittings result in a rather congested spectrum in the microwave range.

We started our analysis of propionamide spectrum from the results of Marstokk et al. (1996) where the data in the 21.4 – 39 GHz range were obtained using Stark modulated spectroscopy. In order to expand the covered frequency range, the new measurements of the propionamide spectrum were performed in three laboratories. Measurements in the range from 9.8 to 25.6 GHz were performed at the National Institute of Standards and Technology (Gaithersburg, USA) using the molecular beam Fourier transform microwave spectrometer (Lovas & Suenram 1987; Suenram et al. 1989). Measurements in the 49 – 149 GHz range were done using the automated millimeter wave spectrometer of the Institute of Radio Astronomy of NASU (Kharkiv, Ukraine) (Alekseev et al. 2012). Measurements between 150 and 460 GHz were done using the submillimeter-wave spectrometer of PhLAM Laboratory at the University of Lille (France) (Zakharenko et al. 2015).

Analysis of the spectra was performed using the Rho axis method (RAM) and the RAM36hf program (Ilyushin et al. 2010; Belloche et al. 2017). The RAM approach allows simultaneous fitting of rotational transitions of A and E symmetry species in a number of methyl torsion excited states. The transitions belonging to the ground ($v_{29}$, $v_{30}$) = (0, 0) and first excited methyl torsion $(v_{29}, v_{30}) = (1, 0)$ states were analyzed simultaneously as the first step of our study. The transitions of the $(v_{29}, v_{30}) = (1, 0)$ state were included mainly with the purpose of better constraining torsional parameters of our RAM Hamiltonian model. In the ground state $(v_{29}, v_{30}) = (0, 0)$ (denoted as v=0), the rotational transitions up to $K_a = 10$, and J up to 50 were included in the fit. Higher $K_a$ (> 10) transitions were excluded from consideration because of quite significant deviation from our model presumably due to intervibrational interactions with low lying skeletal torsion mode $(v_{29}, v_{30}) = (0, 1)$ in this molecule. For the same reason we limited our fitted dataset for the first excited methyl torsion state to $K_a \leq 4$ transitions. Such constraint seems reasonable since lines corresponding to relatively low $K_a$ transitions in the ground vibrational state $(v_{29}, v_{30}) = (0, 0)$ are most likely to be detected in the interstellar medium.

Despite the fact that transitions belonging to the $(v_{29}, v_{30}) = (1, 0)$ state were included in the fit, the main torsional parameters remained highly correlated and we had to fix the F parameter in our RAM Hamiltonian. We fixed the F parameter at 5.55 cm$^{-1}$, a value that corresponds to methyl top moment of inertia $I_\alpha \approx 3.2$ a.m.u. . The final dataset for the $(v_{29}, v_{30}) = (0, 0), (1, 0)$ states contains 5495 transitions which due to blending correspond to 2271 measured line frequencies. This dataset includes our new measurements as well as data for $(v_{29}, v_{30}) = (0, 0)$ from Marstokk et al. (1996). We have not included in the fit the data on 8 rotational transitions from (Marstokk et al. 1996) that correspond to $(v_{29}, v_{30}) = (1, 0)$ state, since those lines deviate significantly from our current model. Whereas we can not rule out completely the presence of some intervibrational interactions for these lines, at present we think that these lines were misassigned in Marstokk et al. (1996). The RAM Hamiltonian model included 26 parameters (with one fixed parameter F ). The final set of molecular parameters is presented in Table 1. The overall root mean square deviation of the fit was 0.035 MHz (corresponding to weighted root mean square deviation of 0.86).

When the first positive results were obtained in our search for interstellar propionamide we decided to extend our analysis to the $(v_{29}, v_{30}) = (0, 1)$ state (denoted as v=1). This state has rather low excitation energy (45(7) cm$^{-1}$; Marstokk et al. (1996)), and thus, it was quite probable that rotational transitions belonging to this state may be also detected in the ISM. As for the ground vibrational state, we performed an analysis for the $(v_{29}, v_{30}) = (0, 1)$ state simultaneously with the corresponding state where one quanta of methyl torsion and one quanta of skeletal torsion were excited $(v_{29}, v_{30}) = (1, 1)$. This was done again to provide more information on the methyl torsion part of the problem. It appeared that excitation of the skeletal torsion vibration increases the barrier to methyl torsion motion (see Table 1 for comparison of $V_3$ values). It is evident from the fact that much smaller A – E splittings are observed for the same rotational transitions in the first excited skeletal torsion state $(v_{29}, v_{30}) = (0, 1)$ in comparison with the ground state $(v_{29}, v_{30}) = (0, 0)$. For example the A – E splitting for the $4_{4,0} \leftarrow 3_{3,1}$ transitions in the ground vibrational state is about 2 MHz, whereas in the first excited skeletal torsion state this splitting is only 0.210 MHz. Thus many transitions for which A – E splittings were resolved in the ground vibrational state $(v_{29}, v_{30}) = (0, 0)$ appeared to be unresolved in the $(v_{29}, v_{30}) = (0, 1)$ state, and therefore, the amount of information about the methyl torsion motion encoded in the A – E splittings of rotational transitions in the first excited skeletal torsion state was even less than in the case of the ground vibrational state. This forced us to fix in the fit for the first excited skeletal torsion state not only the F parameter, but also the ϱ parameter, because if we varied the ϱ parameter the fit provided a value that corresponded to an unphysically low value of methyl top moment of inertia $I_\alpha$. The F and ϱ parameters thus were fixed at the corresponding values from the ground state fit

(see Table 1). The final dataset for the $(v_{29}, v_{30}) = (0, 1), (1, 1)$ fit contains 4017 transitions with J up to 30, which due to blending correspond to 1027 measured line frequencies. The dataset includes our new measurements in the 49 – 146 GHz frequency range as well as data for the $(v_{29}, v_{30}) = (0, 1)$ state from Marstokk et al. (1996). The RAM Hamiltonian model included 25 parameters for the first excited skeletal torsion state (two parameters were fixed). This parameter set is also given in Table 1. The overall root mean square deviation of the fit was 0.046 MHz (corresponding to weighted root mean square deviation of 1.35).

Using the parameters of our final fits, we calculated a list of propionamide transitions in the ground $(v_{29}, v_{30}) = (0, 0)$ and $(v_{29}, v_{30}) = (0, 1)$ states for astronomical use. We used dipole moment components obtained by Marstokk et al. (1996) via Stark measurements $\mu_a = 0.6359(60)$ D and $\mu_b = 3.496(30)$ D, which were rotated from the principal axis system to the rho axis system of our Hamiltonian model. The list includes information on transition quantum numbers, transition frequencies, calculated uncertainties, lower state energies, and transition strengths. We limit our predictions to the transitions with calculated uncertainties lower than 0.1 MHz. It should be noted that calculated uncertainties for transition frequencies can not predict possible systematic errors due to intervibrational interactions between methyl torsion and skeletal torsion modes in the molecule, since those interactions are not taken explicitly into account by our current Hamiltonian model. That is why we further limit our predictions to the rotational quantum number ranges where we do not see in our laboratory study significant deviations from our model. The predictions are made up to 460 GHz and up to J = 40 for the ground vibrational state $(v_{29}, v_{30}) = (0,0)$ and up to 150 GHz and up to J = 30 for the $(v_{29}, v_{30}) = (0,1)$ state. For both states only the transitions with $K_a < 10$ are given to minimize possible errors due to intervibrational interactions between the ground state and the stack of skeletal torsion excited states $v_{30} = 1, 2$. To exclude all doubts in rest frequencies of the propionamide transitions detected in the ISM, we have checked against laboratory experimental records all clean and partially blended lines detected.

Along with transition frequency predictions, we also provide tabulated values of the partition function for propionamide (see Table 2). The rotational $Q_r(T)$ and vibrational $Q_v(T)$ parts of the partition function were calculated using standard expressions:

$$Q_r(T) = g_i g_N \sqrt{\frac{\pi}{ABC}\left(\frac{kT}{h}\right)^3} \quad (1)$$

$$Q_v(T) = \prod_{i=1}^{3N-6} \frac{1}{1 - e^{-E_i/kT}} \quad (2)$$

The total partition function is thus $Q = Q_r Q_v$. In Eq. 1, A, B, and C are rotational constants in the principal axes system, $g_N$ and $g_i$ are degeneracy factors taking into account, respectively, the tripling of energy levels due to nuclear quadrupole coupling of nitrogen, and the doubling of energy levels due to hindered methyl top internal rotation. For the calculation of the vibrational partition function, we used the excitation energies $E_i$ of all normal modes of propionamide up to 1000 cm$^{-1}$. The energies of the two lowest modes $v_{30}$ and $v_{29}$ were taken respectively from the study by Marstokk et al. (1996), and from this study as a result of global modeling of the methyl torsion potential function based on the parameters obtained from RAM36hf fit. The excitation energies of all other modes were taken from the study of the vibrational spectrum of propionamide by Kuroda et al. (1972).

The datasets treated in this work as well as the predictions produced are available as electronic supplement with this paper. Part of them are presented in the Table A1, A2, A3 and A4. Transitions for the ground state $(v_{29}, v_{30}) = (0, 0)$ and the first excited skeletal torsion state (v$_{29}$, v$_{30}$) = (0, 1) are given separately.

## 3. OBSERVATIONS AND DATA REDUCTION

### 3.1. *IRAM 30m telescope*

The observations presented in this paper are part of a mapping observations of molecular lines at 3- and 2-mm wavelength toward Sgr B2 in 2019 May with the Institut de Radioastronomie Millimetrique (IRAM) 30m telescope on Pica Veleta, Spain (project ID: 170-18, PI: Feng Gao). The observing center is Sgr B2(N) ($\alpha_{J2000}$ = 17:47:20.0, $\delta_{J2000}$ = −28:22:16). Observations were conducted in position-switching mode. The off position of ($\delta\alpha$, $\delta\beta$)= (−752″, 342″) was used (Belloche et al. 2013). We use the 3-mm band of the EMIR receiver and FTS backend, with a broad bandpass (7.8 GHz bandwidth), to cover the frequency range of 82.3-90.1 GHz with a frequency resolution of 0.195 MHz (velocity resolution of 0.641 km s$^{-1}$ at 3-mm). The 1$\sigma$ rms T$^*_A$ in the line free channels is 4-8 mK at 3mm. The system temperature ranges from 90 to 150 K. The integration time is 98 minutes for Sgr B2(N). The telescope pointing was checked every ~2 hr on nearby quasar 1757-240. The telescope focus was optimized on 1757-240 at the beginning of the observation. The data reduction was conducted using the GILDAS software package[1], including CLASS, GREG, and Weeds (Maret et al. 2011) (version of April 2020).

In this paper, the intensity is given as the main beam brightness temperature (T$_{mb}$), which is defined as T$_{mb}$ = T$_A^*$ × F$_{eff}$/B$_{eff}$, in which T$_A^*$ is the antenna temperature, F$_{eff}$ is the forward efficiency and B$_{eff}$ is the beam efficiency. The values used were F$_{eff}$ = 0.95 and B$_{eff}$ = 0.81 for 3 mm. The half-power beam widths (HPBW) are ∼ 24″ for observations at 3 mm.

---

[1] http://www.iram.fr/IRAMFR/GILDAS.

## 3.2. ALMA archive data

The interferometric data used here were acquired from the ALMA Science archive of the ReMoCA survey (Re-exploring Molecular Complexity with ALMA) (Belloche et al. 2019). This survey was performed during Cycle 4 between 2016 and 2017. A total of five spectral setups were used to cover frequencies between 84.1 and 114.4 GHz continuously. Each spectral setup was observed independently, repeated by four executions, with each execution achieving an on-source time of 47-50 minutes. The single polarization mode was employed in the spectral setups to achieve a simultaneous handling of four spectral windows of 1.875 GHz bandwidth each, with a uniform spectral resolution of 0.488 MHz that corresponds to a velocity resolution of 1.3-1.7 km s$^{-1}$ across the observing band. The phase center of the observations is between the two hot cores Sgr B2(N1) and (N2), at the coordinates $\alpha_{J2000}$ = 17:47:19.87, $\delta_{J2000}$ = −28:22:16. The HPBW of the primary beam, equivalent to the field of view of the observations, is between 69″ and 51″ from the lowest to the highest observed frequencies. Details about the observations were presented in Belloche et al. (2019).

The data were calibrated using the standard ALMA data calibration pipeline with CASA version 4.7.0-1 for the first spectral setup, and with CASA version 4.7.2 for the other four spectral setups. The calibrated data were then imaged with CASA version 5.6.1-8. The bandpass calibration was carried out on the quasar J1924-2914, with the exception of one execution in the second spectral setup that used the quasar J1517- 2422 instead. The absolute flux density scale was derived from the quasars J1924-2914 or J1733-1304. The phase and amplitude calibrations were performed on the quasar J1744-3116. The images were produced with the TCLEAN deconvolution algorithm in CASA. The HPBW of the synthesized beam depends on the setup and the sideband but is typically $0.6'' \times 0.5''$. The median rms noise level in the channel maps of the data cubes reached for each selected track varies between 0.4 and 1 mJy/beam.

## 4. OBSERVATIONAL RESULTS

### 4.1. *IRAM 30m result*

For the IRAM data we have considered a number of propionamide emission lines among which two lines at 86954 MHz and 87980 MHz looked most promising. Unfortunately our pre-analysis showed that even these two lines suffer from serious line blending from other molecules. As is stated in Section 3.1, the HPBWs are ~ 24" for IRAM observations at 3 mm. Within this range, there are many hot cores, HII regions, protostellar outflows, and so on (Bonfand et al. 2017; Belloche et al. 2019). Thus, the IRAM spectrum of propionamide may include contributions from a number of hot cores in Sgr B2(N), and analysis of propionamide emission in all these components of Sgr B2(N) is needed to model the propionamide emission from Sgr B2(N), which is beyond the scope of this paper. That is why as the next step of our search we concentrated on ALMA archive data (see Section 3.2) which provided the HPBW for a synthesized beam of order of 0.6" × 0.5" thus allowing a search for a position in which propionamide is less contaminated by line blending from other molecules.

### 4.2. *ALMA archive result*

#### 4.2.1. *Position selection*

Sgr B2(N1) is a hot core that is known to be rich in COMs (Belloche et al. 2013, 2019). However, the spectrum toward the continuum peak position of Sgr B2(N1) is severely affected by the optically thick free- free emission from the HII regions, and the weak COM transitions also suffer from serious line blending from other molecules (Belloche et al. 2019). Our pre-analysis indicates that the 92.143 GHz line appears to be the only line that is clean across the whole N1 region. By mapping the integrated intensity of 92.143 GHz line of propionamide and comparing to that of CH$^{13}$CN (Fig. 2), we selected a position in which propionamide is relatively strong and other molecular emissions are relatively weak. The selected region is about 1.5″ to the east of the hot core Sgr B2(N1). The coordinate is 17:47:19.99, -28:22:18.72(J2000). The size of selected region is ∼ 0.3″. This new position, which is studied for the first time, is referred to as Sgr B2(N1E). Thus, as the next step we concentrated on the Sgr B2(N1E) direction to avoid high opacity from strong continuum emission and serious line blending from other molecules.

4.2.2. *Removal of continuum emission.*

The identification and removal of the continuum emission is difficult yet critical for the identification of propionamide in Sgr B2(N). As stated in Belloche et al. (2019), because of the large number of spectral lines, their different systemic velocities, and the velocity gradients, it is impossible to split the line and continuum emission in the Fourier domain. Therefore we performed the splitting in the image plane. As Sgr B2(N1E) is spatially offset from the hot core position, both the continuum emission and molecular line emission toward Sgr B2(N1E) are much weaker than those toward Sgr B2(N1S). The typical linewidth of molecular line emission toward N1E is 4 km s$^{-1}$, which is narrower than the typical linewidth of molecular line emission toward N1S (∼5 km/s, see Table 4 in Belloche et al. (2019). As such, the baseline extraction for N1E is easier than that for N1S. For each spectral window, we selected 3-5 groups of channels that seem to be free of line emission. A first-order baseline was fitted to these channels. The subtracted continuum emission of lines that is obviously offset from the spectrum energy distribution of free-free emission, is regarded as being affected by line confusion or absorption. The subtracted continuum emission for the identified lines are presented in Fig. 3. The overall flux dependence on the frequencies is well consistent with the spectrum energy distribution that is expected by free-free emission.

### 4.2.3. $H_2$ column density

We used the $C^{18}O$ column density to derive the $H_2$ column density. Fig. 4 shows the ALMA spectrum of $C^{18}O$ toward Sgr B2(N1E). The $C^{18}O$ column density has been obtained with WEEDS by assuming optically thin emission and local thermodynamic equilibrium (LTE, which means that the excitation temperature of the transitions is equal to the rotational temperature of the molecule and the kinetic temperature of the gas) conditions. An uniform excitation temperature $T_{ex}$ of 150 K (Belloche et al. 2019) was adopted. The column density of C O was derived to be $2.5 \times 10^{17}$ cm$^{-2}$. According to the Galactic gradient of $^{16}O/^{18}O$(Wilson & Rood 1994), the abundance ratio of $C^{18}O$ is significantly different in the Galactic Center region and the solar neighborhood. We adopt an abundance ratio $N(H_2)/N(C^{18}O)$ two times of that in Orion KL ($7 \times 10^6$ cm$^{-2}$, Castets & Langer (1995)). The $N(H_2)$ was then calculated to be $3.5 \times 10^{24}$ cm$^{-2}$.

### 4.2.4. *Detection of propionamide*

We used Weeds to do line identification and spectral modeling. We built our own Weeds database containing the spectroscopic data of propionamide, acetamide (Ilyushin et al. 2004) and acetone (Ilyushin et al. 2019). The JPL (Pickett et al.1998) and CDMS(Müller et al. 2005; Endresetal.2016) databases are used for other molecules. The 3-mm emission of propionamide toward Sgr B2(N1E) is modeled assuming LTE conditions with five parameters: column density, temperature, source size, velocity offset, and linewidth. We model each spectral window of each observed setup separately to account for the varying angular resolution. A same set of the five parameters is used to model the $C_2H_5CONH_2$ v=0 and v=1 states.

Approximately 110 $C_2H_5CONH_2$ v=0 lines were expected to be higher than 3σ noise level. Six clean lines were detected toward Sgr B2(N1E). These lines are presented in Fig. 5 and Table 3. Approximately 17 partially blended lines were also detected. The remaining 90 lines that match the considered frequency range and noise level threshold are blended with other species, but the observed results do not contradict the expected intensities for these lines. About 40 $C_2H_5CONH_2$ v=1 lines with expected intensities above 3σ noise level were expected. Of these, eight partially blend lines were detected (Fig. 7), with the rest blended seriously with other species. The data behind Figure 5, 6 and 7 are available as electronic supplement with this paper.

With the derived size, a linewidth measurement of 4.2(.5) km s$^{-1}$ with Gaussian fit of 92.408 and 103.342 GHz lines in CLASS, and an assumption of excitation temperature of 150 K (which is very close to what was obtained for same molecules toward Sgr B2(N1S) by Belloche et al. (2019), we simulated the spectral profiles for all transitions of propionamide detected toward Sgr B2(N1E) with Weeds. The emission of the vibrationally excited skeletal torsion state of propionamide was modeled with the same parameters as those for the vibrational ground state. A column density of $1.5 \times 10^{16}$ cm$^{-2}$ was obtained for propionamide (see Table 4). Abundances relative to $H_2$ was estimated to be $4.3 \times 10^{-9}$ for $C_2H_5CONH_2$.

### 4.2.5. *Related molecules*

Fig. 8 and Fig. 9 present transitions of formamide and acetamide that are identified in the ALMA spectrum of Sgr B2(N1E), respectively. Fig. 10 presents integrated intensity maps of $NH_2CHO$ and $CH_3CONH_2$ overlaid on $C_2H_5CONH_2$ toward Sgr B2(N1). The integrated intensity maps of $CH_3CONH_2$ and

NH$_2$CHO indicate that the amide family shows quite similar spatial distributions, implying a consistent formation mechanism in the ISM. To compare the column density derived for C$_2$H$_5$CONH$_2$ to other amide families, the column densities of CH$_3$CONH$_2$ and NH$_2$CHO toward Sgr B2(N1E) were also derived using Weeds package. We derived average abundances relative to H$_2$ of $7.2 \times 10^{-9}$ for CH$_3$CONH$_2$, and $8.0 \times 10^{-8}$ for NH$_2$CHO. The abundance ratio [C$_2$H$_5$CONH$_2$]/[CH$_3$CONH$_2$] was 0.6, while [CH$_3$CONH$_2$]/[NH$_2$CHO] was estimated to be ~0.1 toward Sgr B2(N1E).

Fig. 11 presents the 92.143 GHz transition of C$_2$H$_5$CONH$_2$ that is identified in the ALMA spectrum toward the main region of Sgr B2(N1), while Fig. 12 and Fig. 13 present transitions of NH$_2$CHO and CH$_3$CONH$_2$ that are identified in Sgr B2(N1), respectively. Due to the high column density and large optical depth of NH$_2$CHO toward Sgr B2(N1), some lines are saturated, such as the 102065 MHz, 105465 MHz, and 106108 MHz lines. A fit to the 92.143 GHz line of C$_2$H$_5$CONH$_2$ toward the main Sgr B2(N1) region was also obtained. A mean column density of $3.3 \times 10^{16}$ cm$^{-2}$ was obtained for C$_2$H$_5$CONH$_2$. The column densities derived for CH$_3$CONH$_2$ and NH$_2$CHO are somewhat lower than those derived for Sgr B2(N1S) by Belloche et al. (2019), as these are the average over the entire CH$_3$CONH$_2$ emission region. The abundance ratio [C$_2$H$_5$CONH$_2$]/[CH$_3$CONH$_2$] was estimated to be ~ 0.22, which was lower than that of Sgr B2(N1E). The [CH$_3$CONH$_2$]/[NH$_2$CHO] is ~ 0.16 toward the main Sgr B2(N1) region. Currently the mechanism for the enhancement of propionamide abundance in comparison with acetamide in Sgr B2(N1E) is uncertain.

We have tried to identify as many molecules as we can in the rest of the spectrum with WEEDS. Table 5 lists the identified molecules, as well as the fitted parameters under LTE assumption.

## 5. DISCUSSION

Based on the six clean lines of propionamide toward Sgr B2(N1E), which are consistent with the LTE synthetic spectrum of propionamide, we reported the tentative detection of propionamide in Sgr B2. Future high sensitivity and resolution observations of protostars with narrower line width might help identify more clean transitions and further confirm the detection of this molecule in the ISM. The results presented here provide strong evidence for warm, compact propionamide with high abundance in the hot cores of the Sgr B2 cloud complex. From a chemical perspective, the presence of propionamide is not surprising. As a substituted acetamide and the third member of the amide chemical family, propionamide is likely to have formation mechanisms that are similar to or associated with those of the two smaller amides: formamide and acetamide.

The formation routes of amide molecules in space are not yet well understood (Ligterink et al. 2020). Under the conditions of Sgr B2(N), it has been proposed that acetamide might be formed through ice chemistry since pure gas phase models cannot produce enough acetamide compared with observations (Quan & Herbst 2007; Garrod 2008). Therefore for the next member of amide family, propionamide, we propose three formation routes, all on the ice mantles on cosmic dust particles.

First, propionamide could form from formamide through reaction with ethane (Ligterink et al. 2018). The well-known interstellar formamide was reported to be detected with a relatively large column density toward Sgr B2 (Rubin et al. 1971), thus it is abundant in the source. Ligterink et al. (2018) investigated the products of ice mixtures of methane and isocyanic acid under UV irradiation, and concluded it can lead to formation of propionamide with the aid of ethane as the latter molecule was also found in the products. Therefore we propose that propionamide could be formed via the following reaction:

$$CH_3CH_3 + NH_2CHO \rightarrow CH_3CH_2CONH_2 + 2H. \qquad (3)$$

First, these two reactant molecules, could be dissociated by UV or cosmic ray induced photons to form radicals that are more reactive. These radicals could then more easily form propionamide, as follows:

$$CH_3CH_3 + h\nu \rightarrow CH_3CH_2 + H. \quad (4)$$

$$NH_2CHO + h\nu \rightarrow NH_2CO + H. \quad (5)$$

$$CH_3CH_2 + NH_2CHO \rightarrow CH_3CH_2CONH_2 + H. \quad (6)$$

$$CH_3CH_2 + NH_2CO \rightarrow CH_3CH_2CONH_2. \quad (7)$$

Similarly, propionamide could also be formed from the next member in the amide family: acetamide. Acetamide was detected toward Sgr B2(N) with relatively high abundance (Hollis et al. 2006; Halfen et al. 2011). We propose that acetamide could form propionamide by reacting with methyl radical directly or via the route of losing one hydrogen atom to form a radical itself and then reacting with the methyl radical, as shown below:

$$CH_3 + CH_3CONH_2 \rightarrow CH_3CH_2CONH_2 + H. \quad (8)$$

$$CH_3CONH_2 + h\nu \rightarrow CH_2CONH_2 + H. \quad (9)$$

$$CH_3 + CH_2CONH_2 \rightarrow CH_3CH_2CONH_2. \quad (10)$$

The third route to form propionamide on dust grains is from isocyanic acid. Garrod (2008) generated possible acetamide by adding methyl radicals to isocyanic acid followed by a hydrogenation step. We propose a similar route could form propionamide as well:

$$HNCO + CH_3CH_2 \rightarrow CH_3CH_2CONH. \quad (11)$$

$$CH_3CH_2CONH + H \rightarrow CH_3CH_2CONH_2. \quad (12)$$

The enhancement of propionamide abundance in comparison with acetamide toward Sgr B2(N1E) is unclear. The physical or chemical conditions in Sgr B2(N1E) might be favored for production or desorption of propionamide. For example, on the grain-surface or ice-mantle in Sgr B2(N1E), there might be abundant methyl radicals, which could react with acetamide and released to gas-phase via some desorption mechanisms. Future detailed chemical modeling and laboratory simulation will help investigate this issue, as well as the formation route that dominants the formation of propionamide.

6. SUMMARY

The high sensitivity and high angular resolution of the ReMoCA spectral line survey conducted with ALMA toward Sgr B2(N) provides a good opportunity for the detection of new complex organic molecules in space. In this paper, using the ReMoCA survey, we present the first tentative interstellar detection of propionamide, $C_2H_5CONH_2$, the next member of the interstellar amide chemical family after formamide and acetamide. As a preparatory step, a new laboratory spectroscopic study of the propionamide spectrum was performed in the 9-461 GHz. The rho axis method and RAM36hf program were used to analyze the propionamide spectrum. By mapping propionamide versus $CH^{13}CN$, we selected a position $1.5''$ to the east of the hot core Sgr B2(N1) (referred to here as Sgr B2(N1E)) where propionamide is relatively strong while other molecular emissions are relatively weak.

We report the tentative detection of propionamide toward Sgr B2(N1E) with six clean lines in the ground vibrational state. More than 20 partially blended lines further support our ISM detection. Our results indicate that the spatial distribution of propionamide is similar to that of formamide and acetamide, suggesting similar formation routes for the amide family. The abundance ratio $[C_2H_5CONH_2]/[CH_3CONH_2]$ toward Sgr B2(N1E) is 0.6, while toward Sgr B2(N1) is estimated to be ~ 0.22. The enhancement of propionamide abundance in comparison with acetamide toward Sgr B2(N1E) is unclear.

The detection of propionamide in Sgr B2(N) demonstrates that interstellar chemistry can reach sufficient levels of complexity to form relatively large peptide molecules and shows the possible growth of larger amide molecules from smaller ones in a massive star-forming process. It is probable that propionamide might also exist in massive star-forming regions in the Galactic disk, such as Orion KL and NGC 6334 ([Ligterink et al. 2020](#)).

## Acknowledgements

The authors thank the staff at IRAM for their excellent support of these observations. We thank S. Bardeau, J. Pety, V. de Souza for help with the Gildas Weeds package. This work made use of the CDMS and JPL Database. J.L. would like to thank discussion with Dr. Liang Jing and Prof. Zhiyu Zhang. This work has been supported in part by the Natural Science Foundation of China (11773054, U1731237, 11590780, 11590784, 11973075, and U1931104). Q.G. acknowledges support of the Fundamental Research Funds for the Central Universities (Grant No. 2020CDJXZ002). RAM/LM acknowledge the support of the Programme National "Physique et Chimie du Milieu Interstellaire" (PCMI) of CNRS/INSU. FJL thanks NIST for providing a laboratory and spectrometer to make some spectral measurements. This paper makes use of the following ALMA data: ADS/JAO.ALMA#2016.1.00074.S. ALMA is a partnership of ESO (representing its member states), NSF (USA), and NINS (Japan), together with NRC (Canada) and NSC and ASIAA (Taiwan), in cooperation with the Republic of Chile. The Joint ALMA Observatory is operated by ESO, AUI/NRAO, and NAOJ. The single dish data are available in the IRAM archive at https://www.iram- institute.org/EN/content-page-386-7-386-0-0-0.html. The interferometric data are available in the ALMA archive at https://almascience.eso.org/aq/.

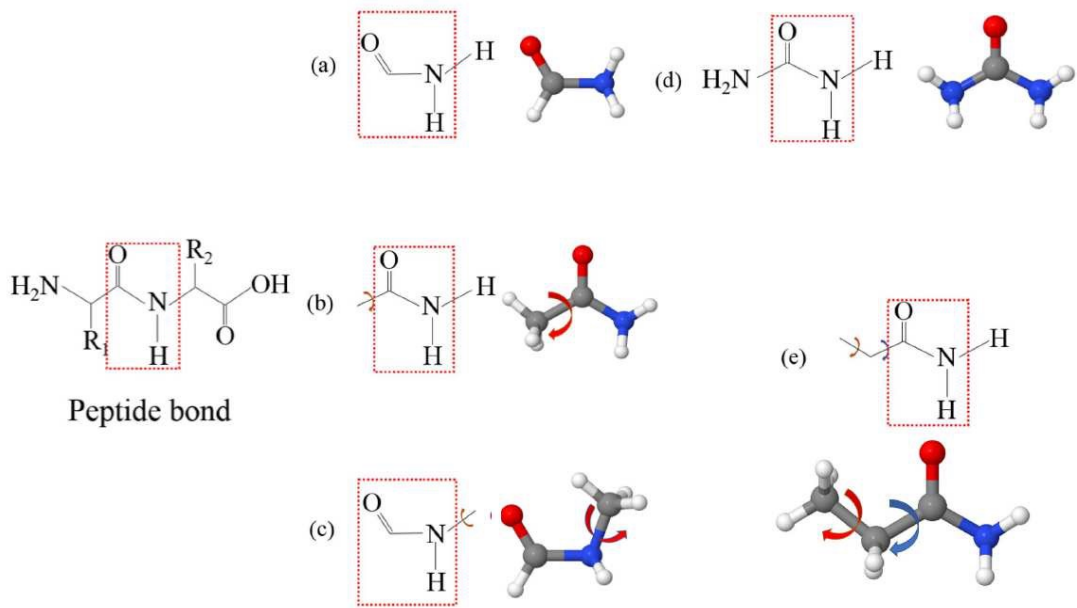

Figure 1. Generalized structures and models of formamide, acetamide, N-methylformamide, urea, and propionamide. (a) $NH_2CHO$. (b) $CH_3CONH_2$. (c) $CH_3NHCHO$. (d) $NH_2CONH_2$. (e) $C_2H_5CONH_2$. The far left panel shows a peptide-bond connecting two amino acids, where R1 and R2 are so-called side-chains. Peptide-bonds are highlighted with red dashed squares. Red arrow denotes hindered internal rotation (torsion) of methyl group; blue arrow stands for skeletal torsion motion around corresponding C-C bond. C, H, N, and O atoms are indicated by gray, small light gray, blue, and red spheres, respectively.

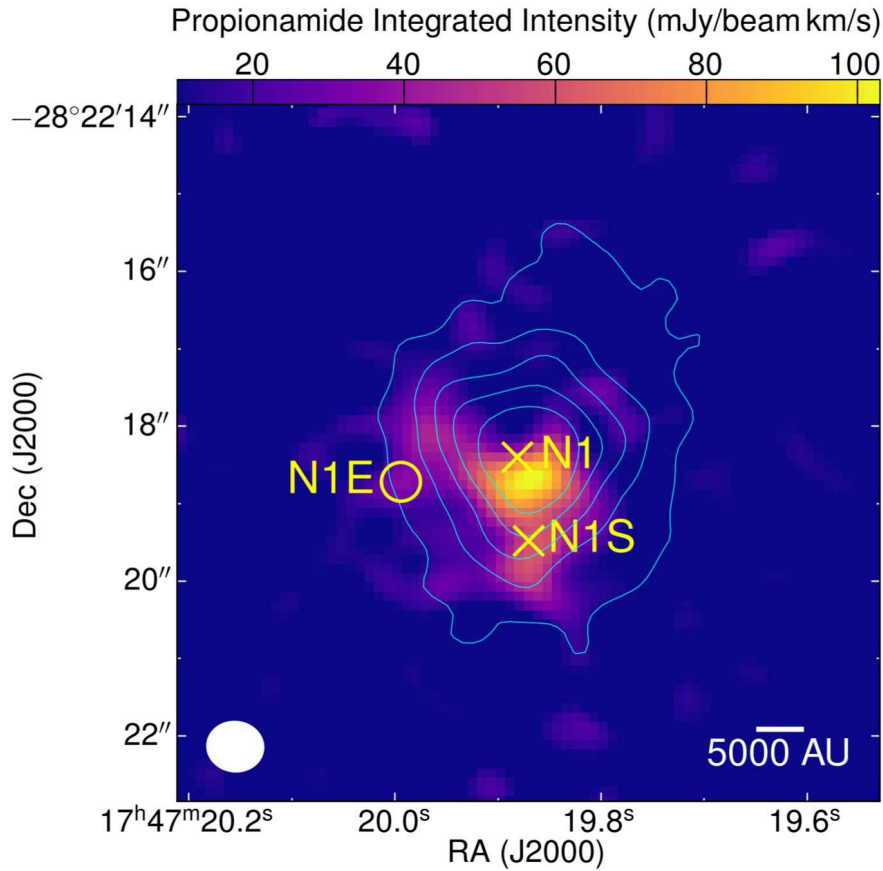

Figure 2. Integrated intensity map of $C_2H_5CONH_2$ $5_{5,*}-4_{4,*}$ transition ($E_u$ = 12.64 K) at 92.143 GHz in Sgr B2(N1) (the $J_{K_a,*}$ notation stands for the two levels with the same $J$, $K_a$ but different $K_c$ quantum numbers ($K_c=J-K_a$ and $K_c=J-K_a+1$) which are degenerate). The integrated intensity map of $C_2H_5CONH_2$ are shown in color scale. The integrated intensity map of $CH^{13}CN$ is shown in contours, which represent 1, 3, 5, 7, 9, 11, 15, 19, 23, 27, 31, 35 Jy/beam. The yellow circle indicates the position of Sgr B2(N1E). The white ellipse shows the size of the respective synthetic beam. The crosses indicate the position of Sgr B2(N1) and Sgr B2(N1S). The integrated frequency range of $C_2H_5CONH_2$ ranges from 92142 to 92143.5 MHz.

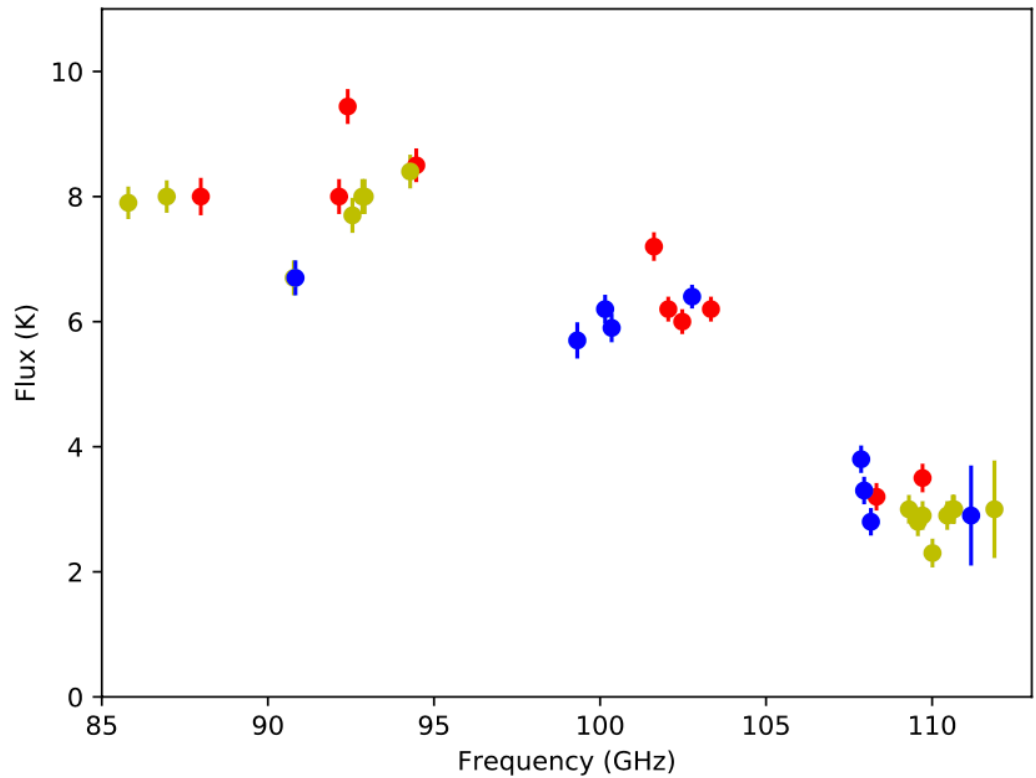

Figure 3. Fluxes of the subtracted continuum emission for the identified lines toward Sgr B2(N1E). The red circles denote fluxes subtracted from the v=0 clean lines. The blue and yellow circles demote fluxes subtracted from the v=0 and v=1 partially blended lines, respectively. The errorbars stand for uncertainties of fluxes. The overall flux dependence on the frequencies is well consistent with the spectrum energy distribution that is expected by free-free emission, which justifies our continuum removal results.

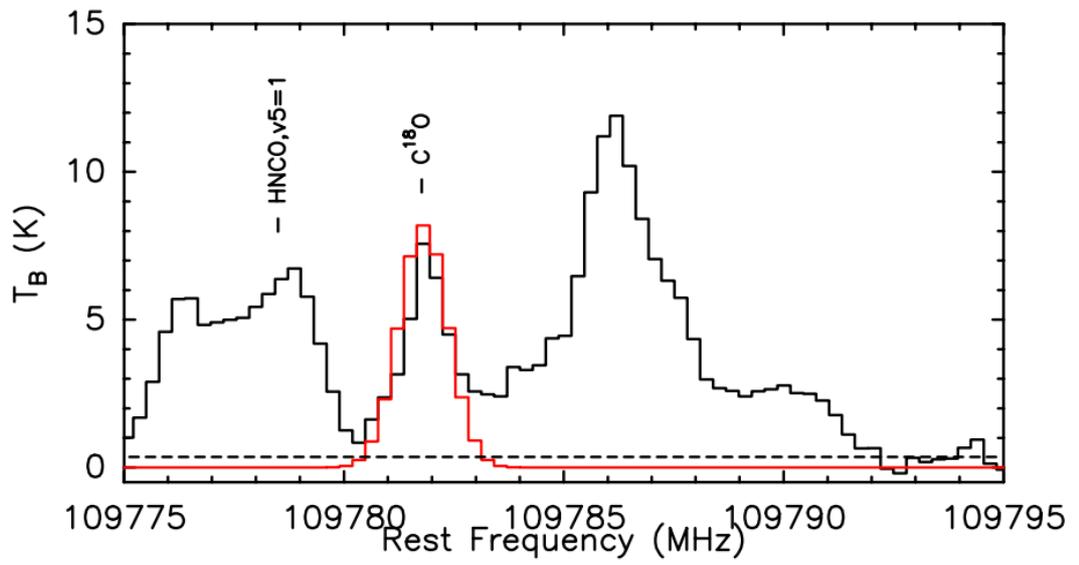

Figure 4. Transitions of $C^{18}O$ that are identified in the ALMA spectrum of Sgr B2(N1E). Black lines show continuum-subtracted spectrum observed with the ALMA telescope, while red lines show the synthetic spectra of $C^{18}O$. The dashed lines denote the $3\sigma$ levels.

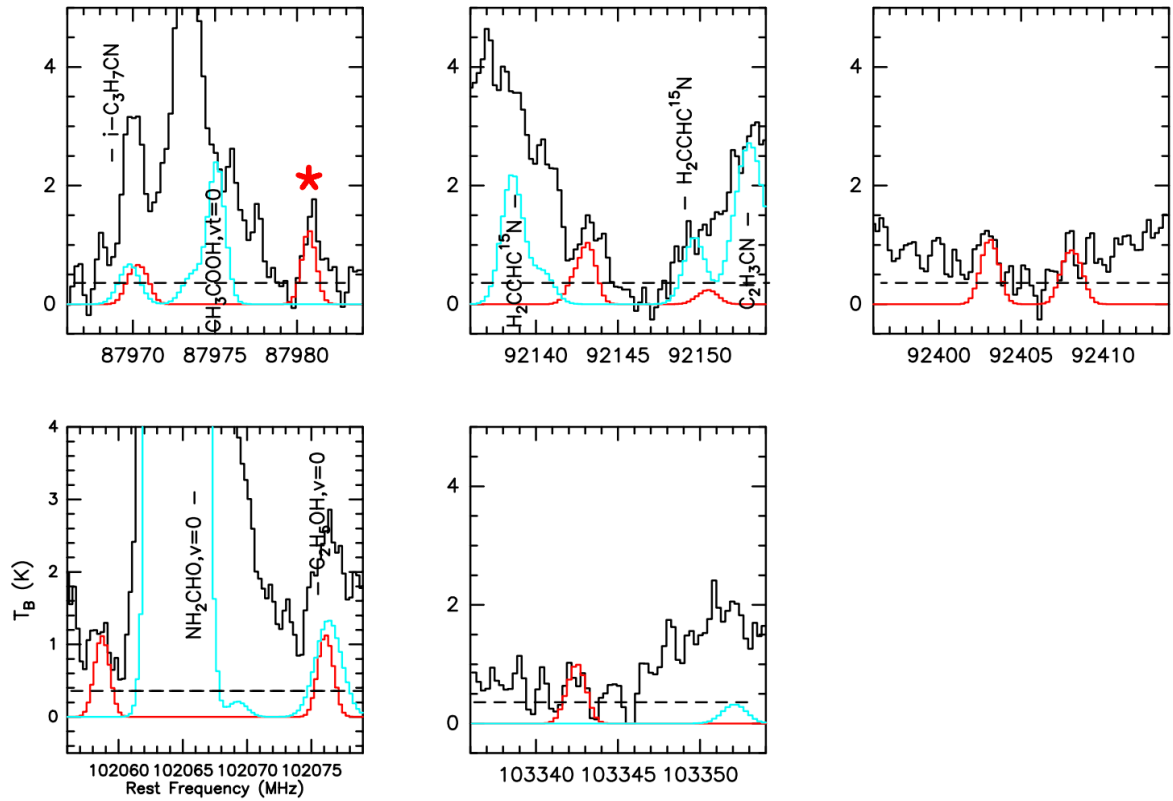

Figure 5. Clean transitions of $C_2H_5CONH_2$ v=0 toward Sgr B2(N1E). Continuum-subtracted spectrum observed with ALMA in black, the synthetic spectra of $C_2H_5CONH_2$ in red, and the preliminary model of other identified molecules in cyan. Red star denotes the 87980 MHz line that has been detected with IRAM 30m. The dashed lines denote the 3σ levels (σ ranges from 0.12 to 0.4 K).

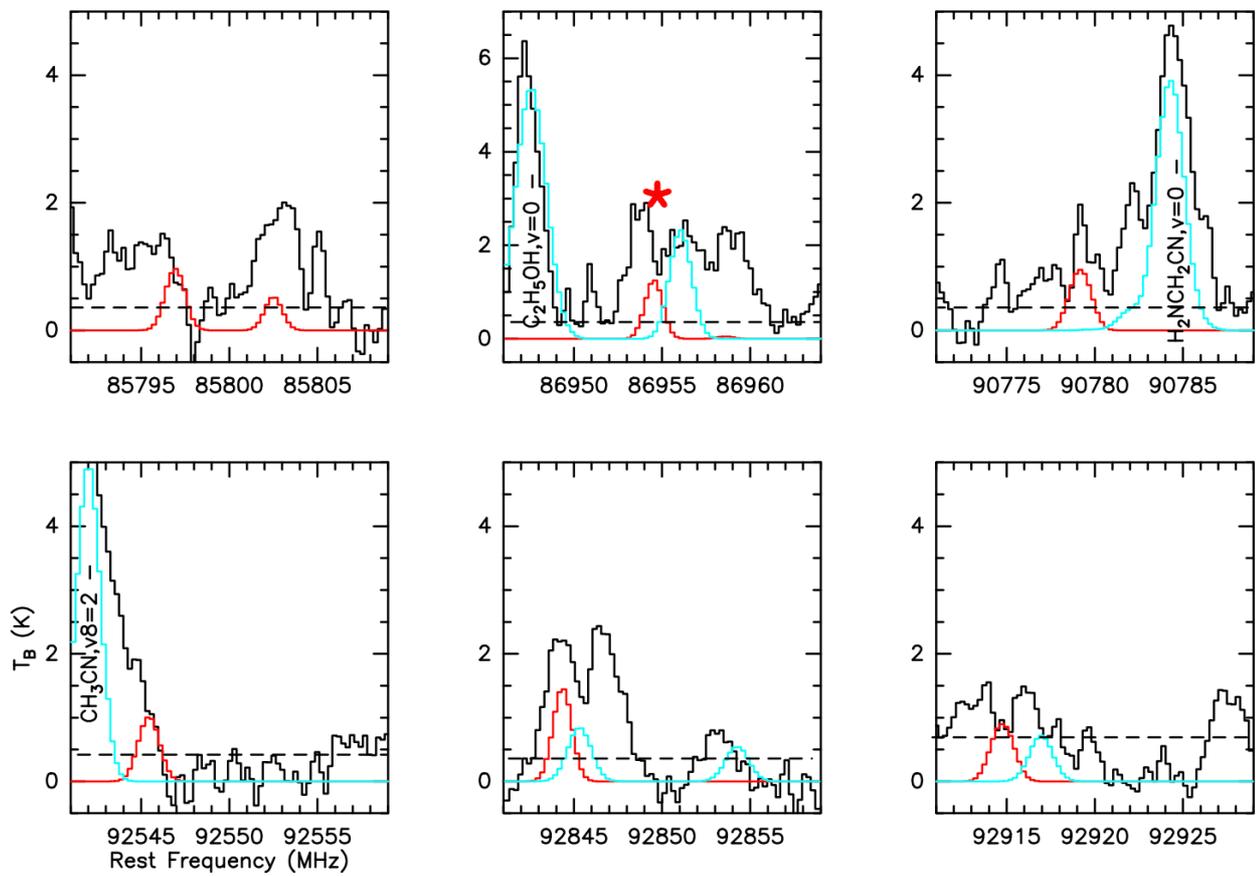

Figure 6. Partially blended transitions of $C_2H_5CONH_2$ v=0 detected with the ALMA telescope toward the Sgr B2(N1E). See caption of Fig. 5 for more detail.

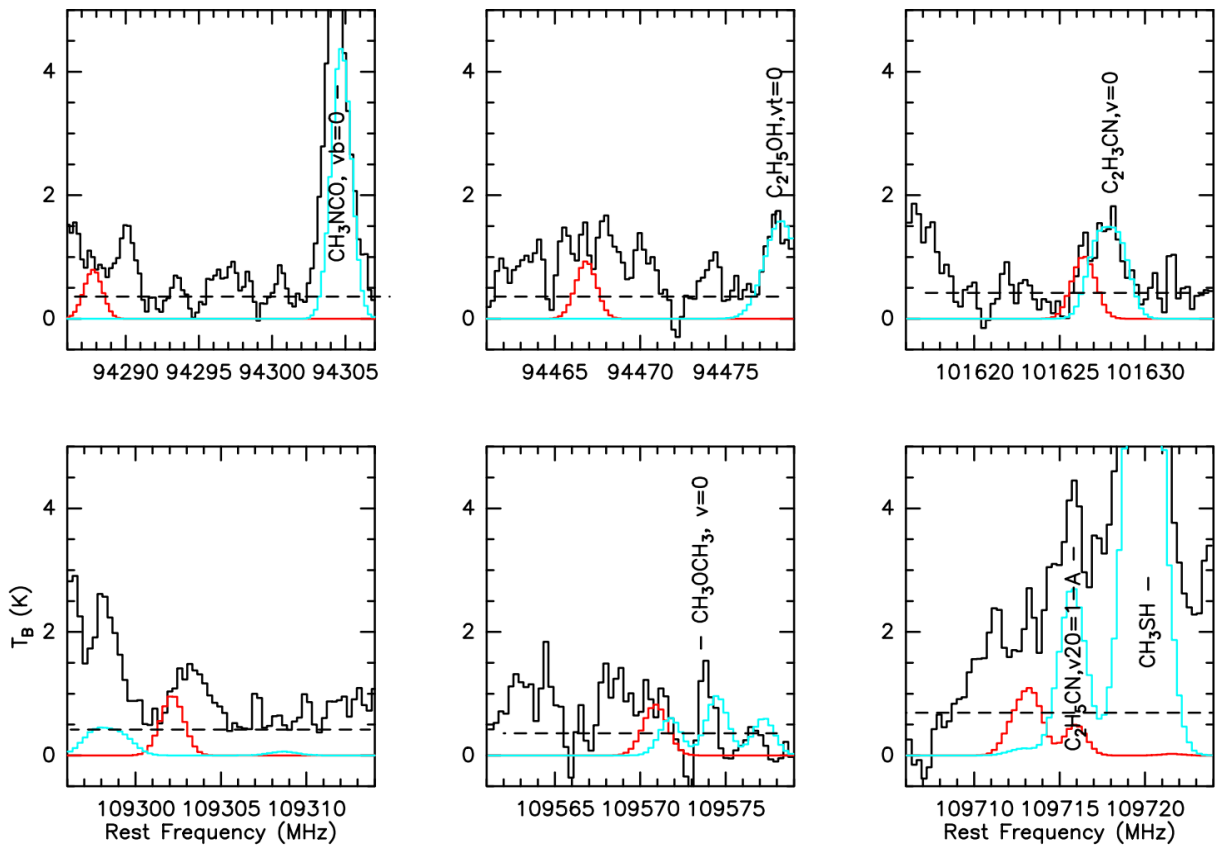

Figure 6. Continued.

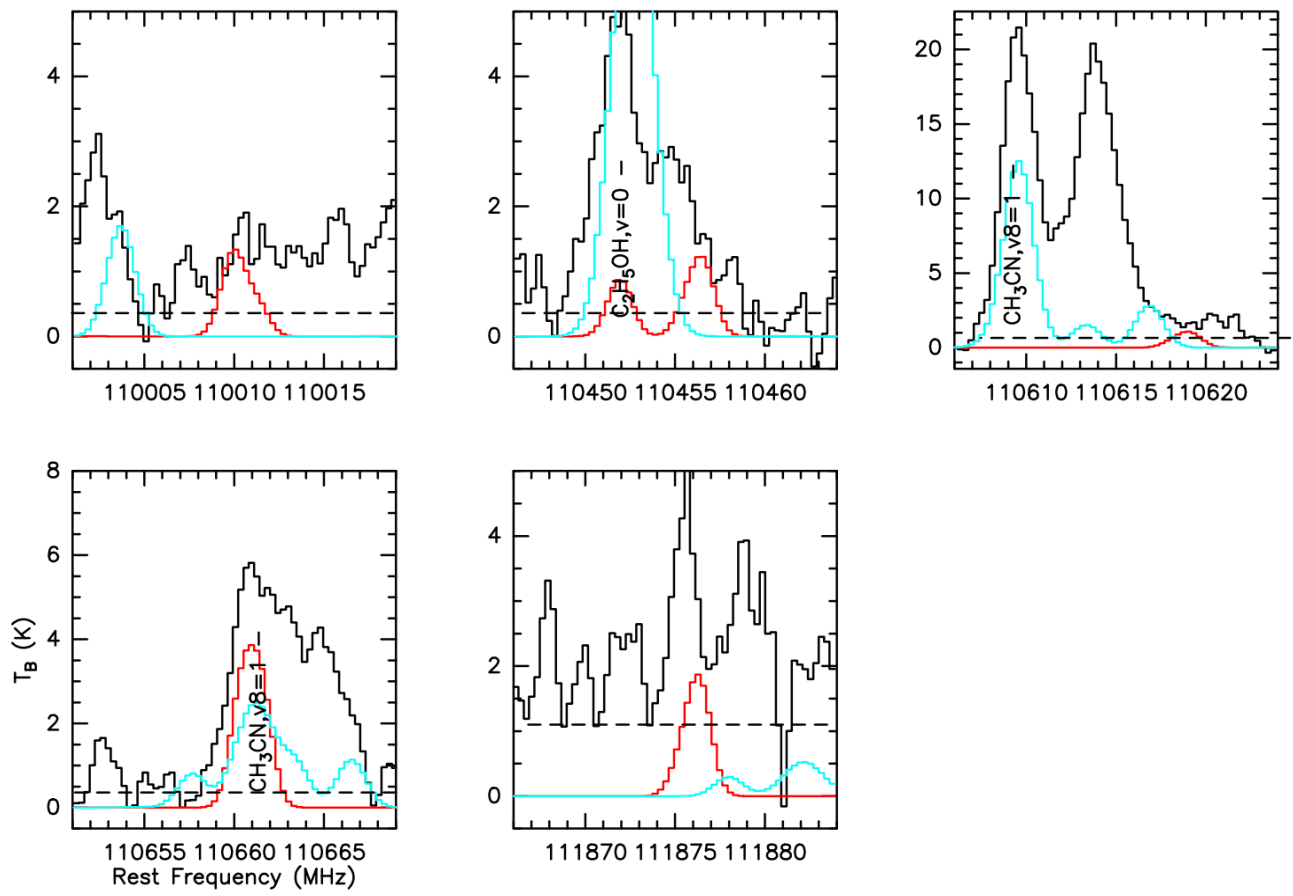

Figure 6. Continued.

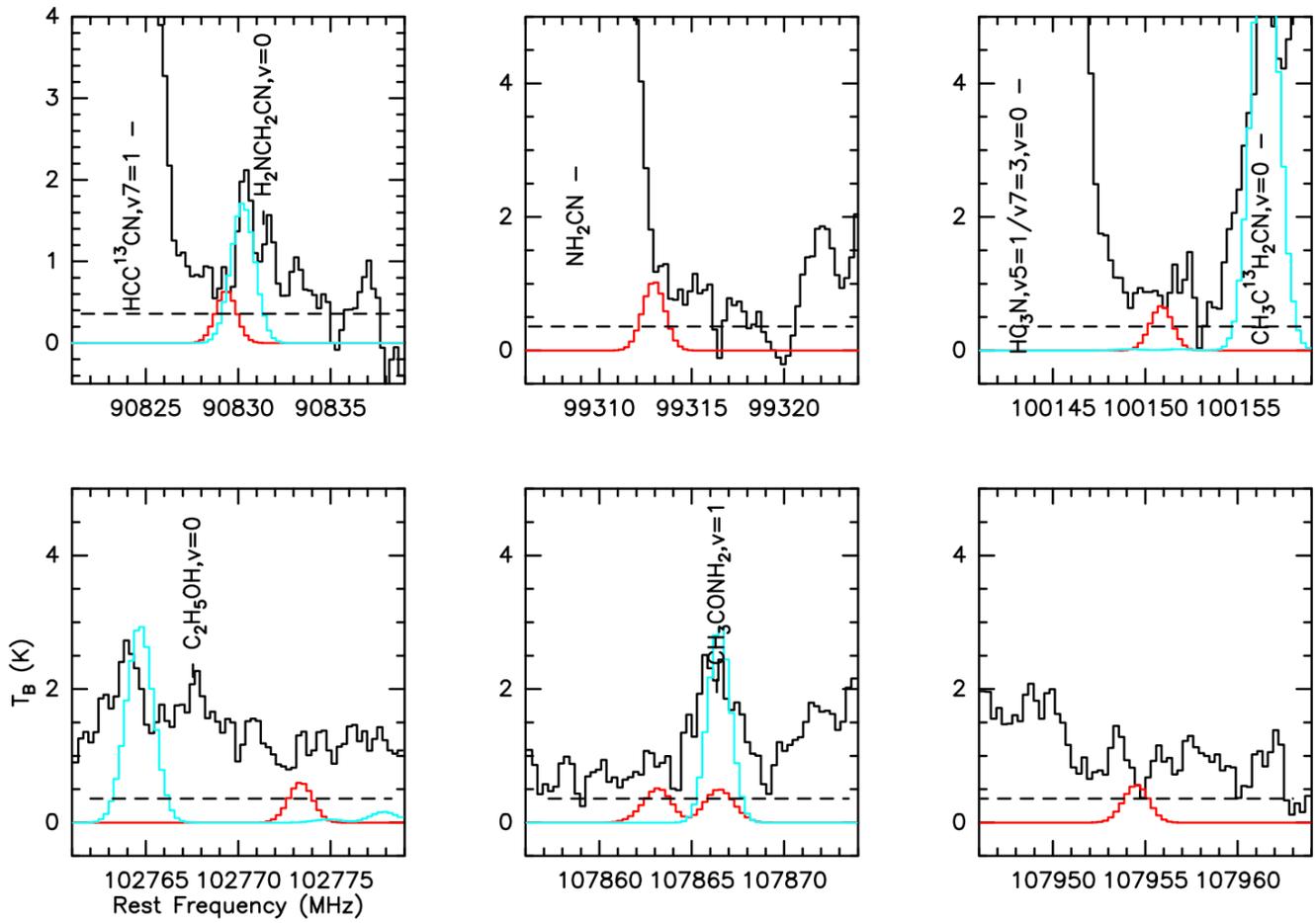

Figure 7. Partially blended transitions of $C_2H_5CONH_2$ v=1 detected with the ALMA telescope toward the Sgr B2(N1E). See caption of Fig. 5 for more detail.

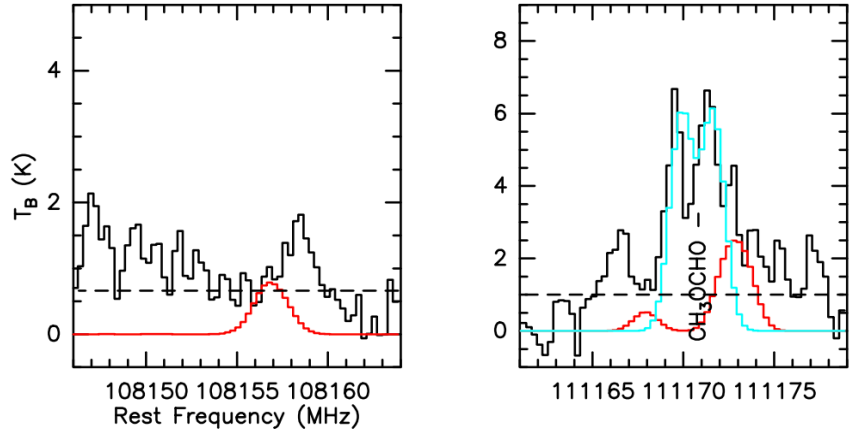

Figure 7. Continued.

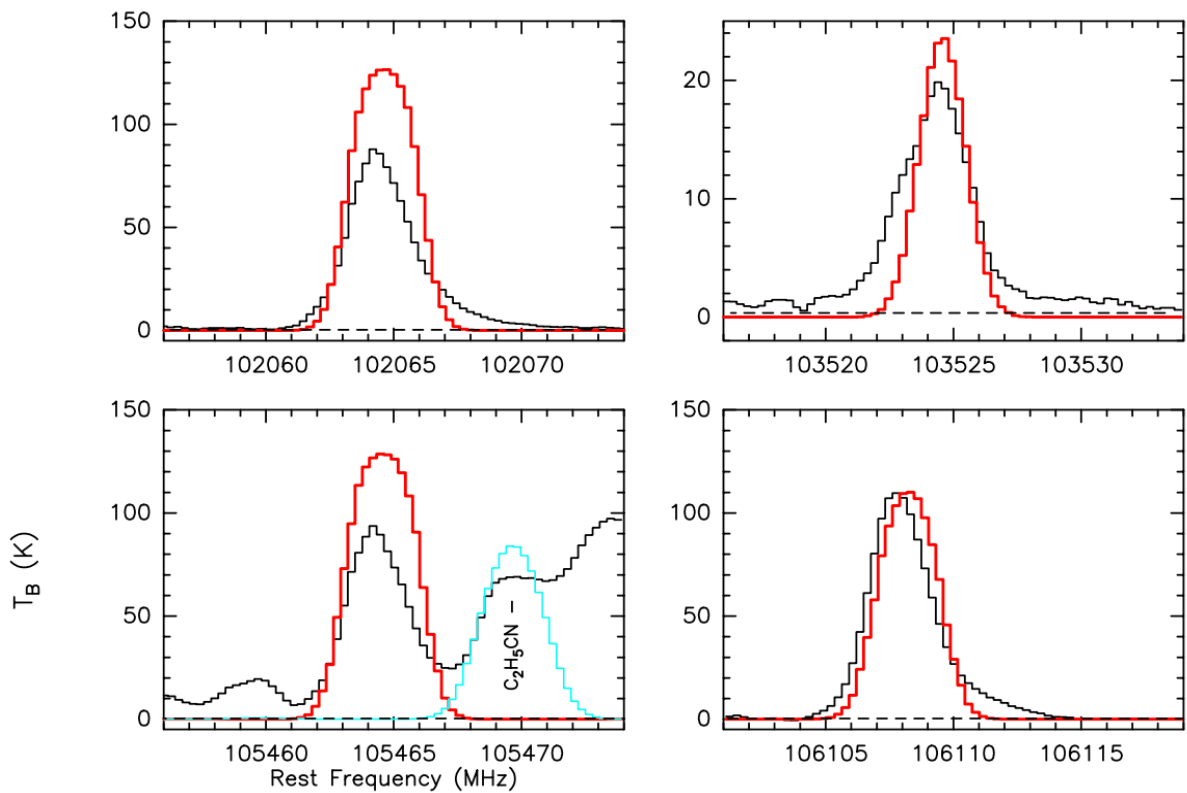

Figure 8. Transitions of NH$_2$CHO that are identified in the ALMA spectrum of Sgr B2(N1E). Black lines show continuum-subtracted spectrum observed with the ALMA telescope, while red lines show the synthetic spectra of NH$_2$CHO. The dashed lines denote the 3σ levels. Obvious discrepancies between synthetic and observed spectra are caused by saturation effects.

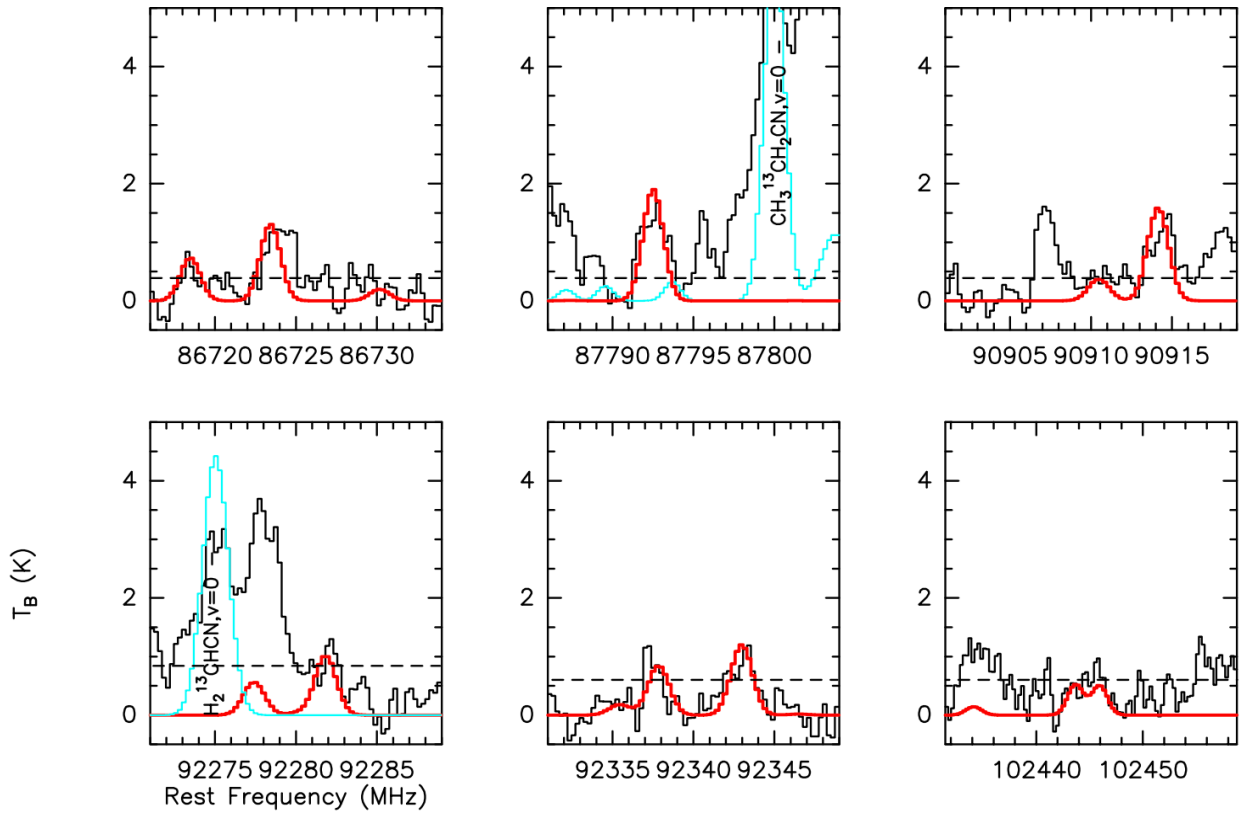

Figure 9. Transitions of $CH_3CONH_2$ that are identified in the ALMA spectrum of Sgr B2(N1E). Black lines show continuum-subtracted spectrum observed with the ALMA telescope, while red lines show the synthetic spectra of $CH_3CONH_2$. The dashed lines denote the $3\sigma$ levels.

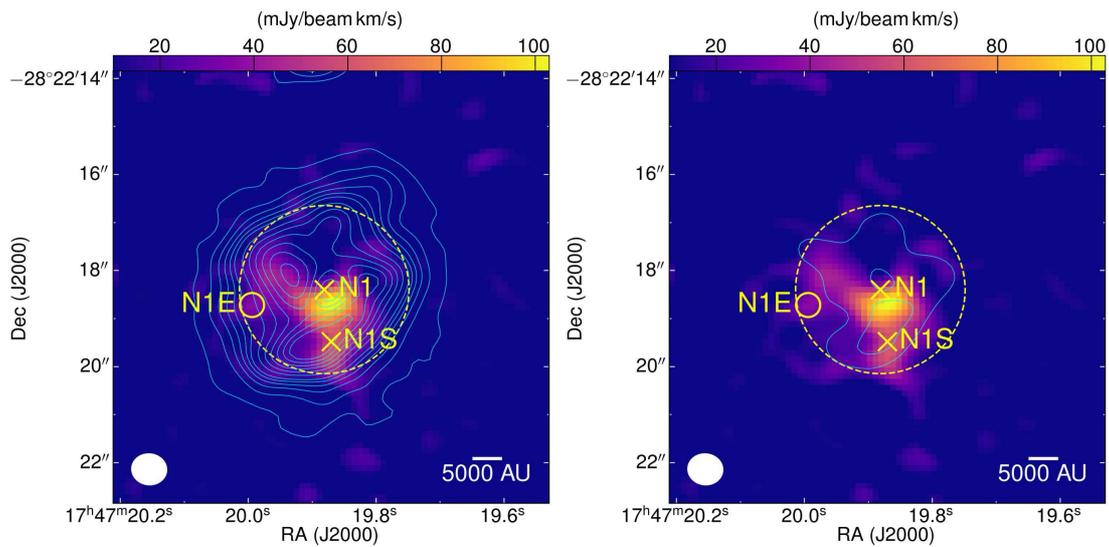

Figure 10. Integrated intensity maps of NH$_2$CHO 103.525 GHz line (left) and CH$_3$CONH$_2$ 102.445 GHz line (right) overlaid on integrated intensity map of C$_2$H$_5$CONH$_2$ 92.143 GHz line in Sgr B2(N1). The contour maps show the integrated intensity maps of NH$_2$CHO and CH$_3$CONH$_2$. The contours represent 5, 15, 25, 35, 45, 55, 75, 95, 115, 135, 155, 175 times of σ, with σ the rms noise level equal to 0.01 Jy beam km/s. The yellow solid circles indicate the position of Sgr B2(N1E), while the yellow dashed circles indicate the range of the selected main Sgr B2(N1) region. The white ellipses show the size of the respective synthetic beams. The crosses indicate the positions of Sgr B2(N1) and Sgr B2(N1S). The integrated frequency range of NH$_2$CHO ranges from 103521 to 103527 GHz, while the integrated frequency range of CH$_3$CONH$_2$ ranges from 102440 to 102448 GHz.

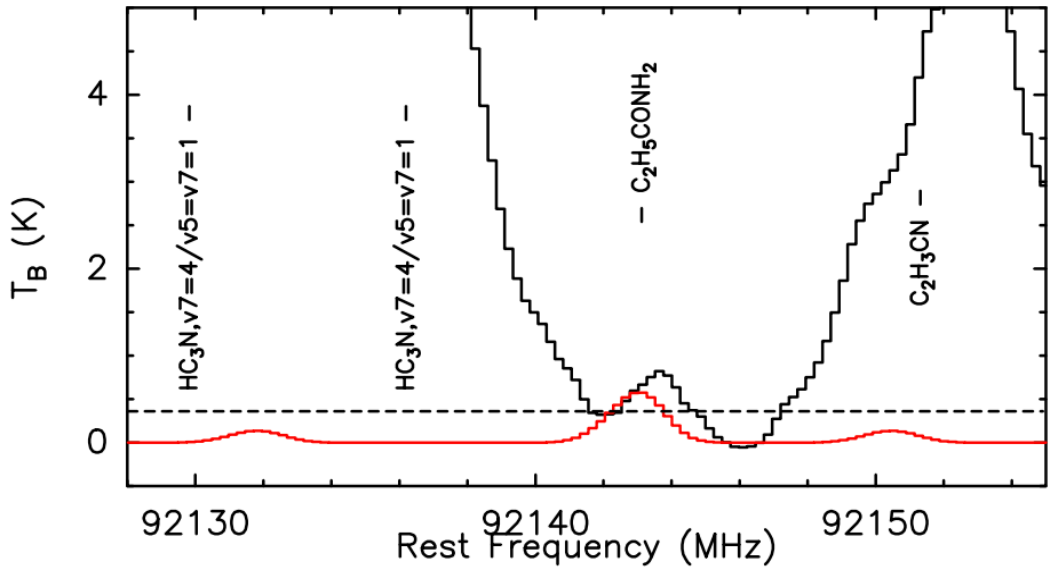

Figure 11. The 92.143 GHz transition of $C_2H_5CONH_2$ that is identified in the ALMA spectrum of Sgr B2(N1). Black lines show continuum-subtracted spectrum observed with the ALMA telescope, while red lines show the synthetic spectra of $C_2H_5CONH_2$. The dashed lines denote the 3σ levels.

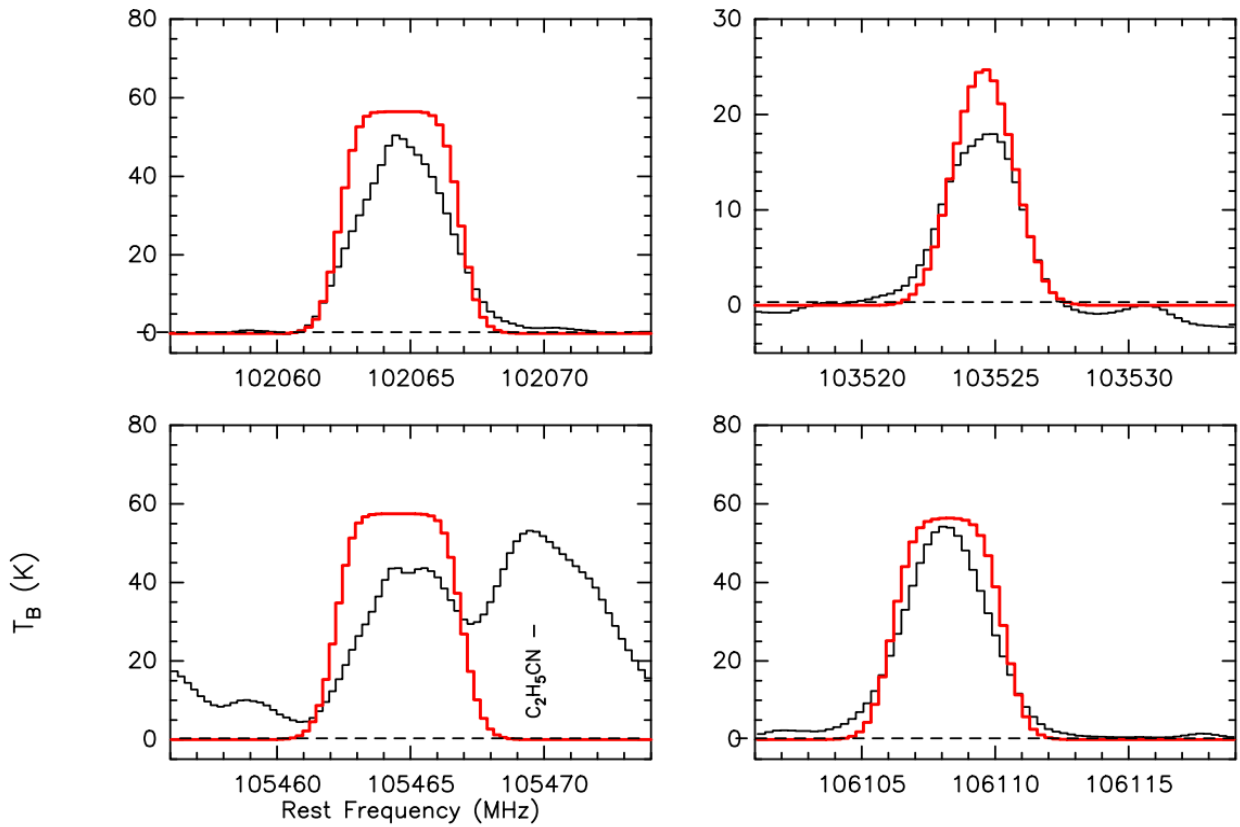

Figure 12. Transitions of NH$_2$CHO that are identified in the ALMA spectrum of Sgr B2(N1). Black lines show continuum-subtracted spectrum observed with the ALMA telescope, while red lines show the synthetic spectra of NH$_2$CHO. The dashed lines denote the 3σ levels. Obvious discrepancies between synthetic and observed spectra are caused by saturation effects.

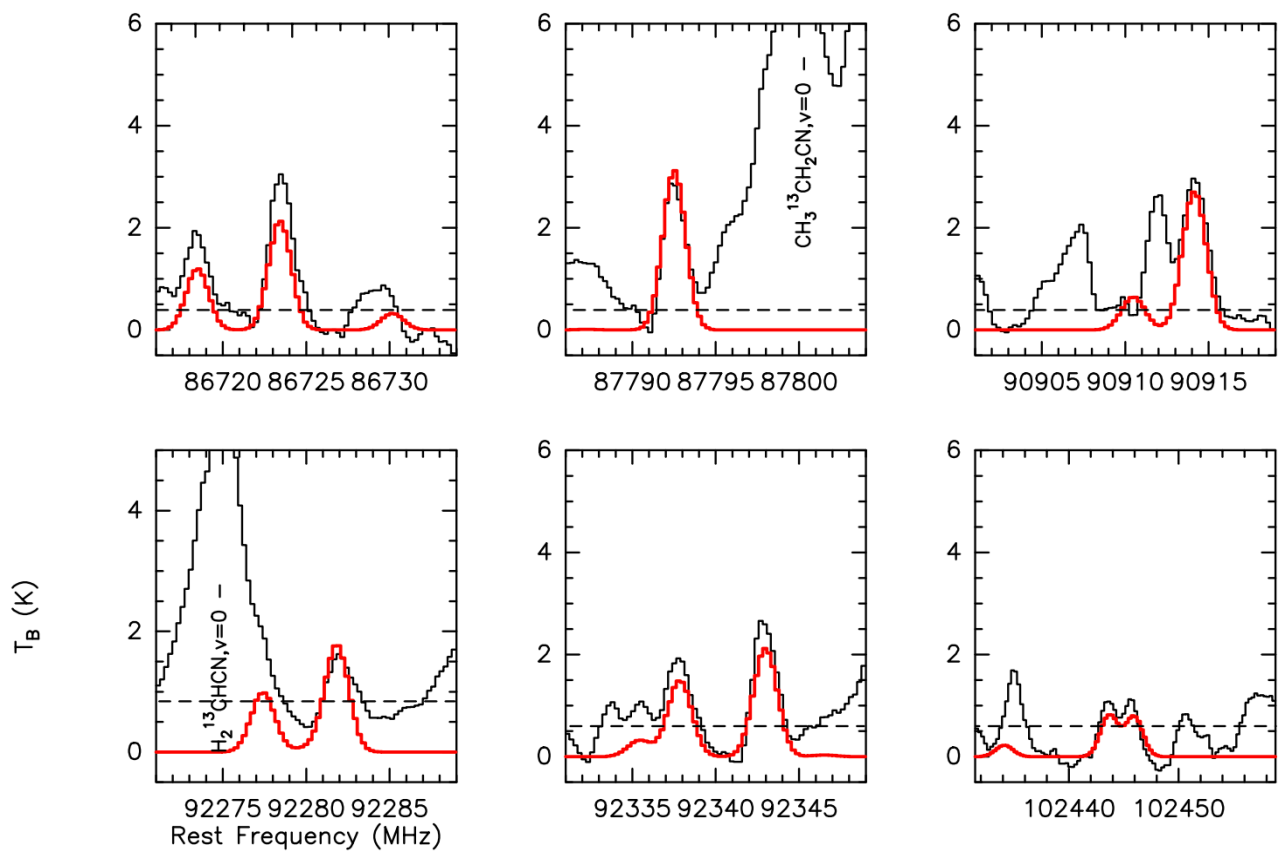

Figure 13. Transitions of $CH_3CONH_2$ that are identified in the ALMA spectrum of Sgr B2(N1). Black lines show continuum-subtracted spectrum observed with the ALMA telescope, while red lines show the synthetic spectra of $CH_3CONH_2$. The dashed lines denote the 3σ levels.

Table1. Molecular parameters of propionamide obtained with the RAM36hf program.

| $ntr^a$ | Parameter | Operator[b] | $v = 0^c$ | $v = 1^c$ |
|---|---|---|---|---|
| 220 | $F$ | $p_\alpha^2$ | 5.55[d] | 5.55[d] |
| 220 | $V_3$ | $\frac{1}{2}(1 - \cos 3\alpha)$ | 759.46(16) | 1042.8(14) |
| 211 | $\rho$ | $J_z p_\alpha$ | 0.054475(70) | 0.054475[d] |
| 202 | $A_{RAM}$ | $J_z^2$ | 0.317819(34) | 0.31562(33) |
| 202 | $B_{RAM}$ | $J_x^2$ | 0.138258(34) | 0.13829(31) |
| 202 | $C_{RAM}$ | $J_y^2$ | 0.0949374576(41) | 0.096130(29) |
| 202 | $D_{ab}$ | $\{J_z, J_x\}$ | $-0.047201(65)$ | $-0.04556(63)$ |
| 440 | $F_m$ | $p_\alpha^4$ | -- | $-0.3270(50) \times 10^{-2}$ |
| 431 | $\rho_m$ | $J_z p_\alpha^3$ | -- | $0.499(21) \times 10^{-3}$ |
| 422 | $F_J$ | $J^2 p_\alpha^2$ | -- | $0.844(30) \times 10^{-4}$ |
| 422 | $V_{3J}$ | $J^2(1 - \cos 3\alpha)$ | $-0.59516(20) \times 10^{-3}$ | $0.234(10) \times 10^{-1}$ |
| 422 | $V_{3K}$ | $J_z^2(1 - \cos 3\alpha)$ | $-0.191946(67) \times 10^{-2}$ | $-0.341(28) \times 10^{-2}$ |
| 422 | $V_{3bc}$ | $(J_x^2 - J_y^2)(1 - \cos 3\alpha)$ | $-0.14119(18) \times 10^{-3}$ | $-0.1125(15) \times 10^{-3}$ |
| 404 | $\Delta_J$ | $-J^4$ | $0.28990(20) \times 10^{-7}$ | $0.128(12) \times 10^{-7}$ |
| 404 | $\Delta_{JK}$ | $-J^2 J_z^2$ | $0.4182(37) \times 10^{-7}$ | $0.299(11) \times 10^{-6}$ |
| 404 | $\Delta_K$ | $-J_z^4$ | $0.30071(53) \times 10^{-6}$ | $0.4772(67) \times 10^{-6}$ |
| 404 | $\delta_J$ | $-2\{J^2, (J_x^2 - J_y^2)\}$ | $0.76905(88) \times 10^{-8}$ | $0.5872(90) \times 10^{-8}$ |
| 404 | $\delta_K$ | $-\{J_z^2, (J_x^2 - J_y^2)\}$ | $-0.100211(26) \times 10^{-6}$ | $-0.8083(73) \times 10^{-7}$ |
| 642 | $D_{6bc}$ | $\frac{1}{2} \sin 6\alpha \{J_x, J_y\}$ | -- | $-0.1767(33) \times 10^{-2}$ |
| 642 | $V_{6J}$ | $J^2(1 - \cos 6\alpha)$ | -- | $-0.1135(47) \times 10^{-1}$ |
| 624 | $V_{3JJ}$ | $J^4(1 - \cos 3\alpha)$ | $0.545(31) \times 10^{-9}$ | -- |
| 606 | $\Phi_J$ | $J^6$ | $0.5310(38) \times 10^{-12}$ | -- |
| 606 | $\Phi_{JK}$ | $J^4 J_z^2$ | $-0.12542(88) \times 10^{-10}$ | $0.597(25) \times 10^{-11}$ |
| 606 | $\Phi_{KJ}$ | $J^2 J_z^4$ | $0.1464(47) \times 10^{-10}$ | $-0.3293(52) \times 10^{-10}$ |
| 606 | $\phi_J$ | $2J^4(J_x^2 - J_y^2)$ | $0.2536(19) \times 10^{-12}$ | -- |
| 606 | $\phi_{JK}$ | $J^2\{J_z^2, (J_x^2 - J_y^2)\}$ | $-0.3479(28) \times 10^{-11}$ | -- |
| 606 | $\phi_K$ | $\{J_z^4, (J_x^2 - J_y^2)\}$ | $0.453(19) \times 10^{-11}$ | $-0.533(33) \times 10^{-11}$ |
| 808 | $L_{KKJ}$ | $J^2 J_z^6$ | $-0.2497(57) \times 10^{-13}$ | -- |
| 808 | $l_K$ | $\{J_z^6, (J_x^2 - J_y^2)\}$ | $0.1454(58) \times 10^{-13}$ | -- |
|  | $\chi_{aa}$ |  | $0.67790(56) \times 10^{-4}$ | $0.639(30) \times 10^{-4}$ |
|  | $\chi_{bb}$ |  | $0.64887(60) \times 10^{-4}$ | $0.644(17) \times 10^{-4}$ |

[a] $n = t + r$, where $n$ is the total order of the operator, $t$ is the order of the torsional part and $r$ is the order of the rotational part, respectively. [b] $\{A, B\} = AB + BA$. The product of the operator in the third column of a given row and the parameter in the second column of that row gives the term actually used in the torsion-rotation Hamiltonian of the program, except for $F$, $\rho$ and $A_{RAM}$, which occur in the Hamiltonian in the form $F(p_\alpha - \rho P_a)^2 + A_{RAM} P_a^2$. [c] All values are in cm$^{-1}$ (except $\rho$ which is unitless). Statistical uncertainties are shown as one standard uncertainty in the units of the last two digits. $v = 0$ correspond to $(v_{29}, v_{30}) = (0, 0), (1, 0)$ dataset, $v = 1$ correspond to $(v_{29}, v_{30}) = (0, 1), (1, 1)$ dataset. [d] Fixed value.

Table 2. Rotational and vibrational partition functions of propionamide at various temperatures.

| $T$ (K) | $Q_r(T)$ | $Q_v(T)$ |
|---|---|---|
| 300.000 | 509564.2 | 21.422 |
| 225.000 | 330971.7 | 8.915 |
| 150.000 | 180158.2 | 3.903 |
| 75.000 | 63695.5 | 1.796 |
| 37.500 | 22519.8 | 1.218 |
| 18.750 | 7961.9 | 1.033 |
| 9.375 | 2815.0 | 1.001 |

Table 3. Rest frequencies and transition parameters of clean C$_2$H$_5$CONH$_2$ lines detected toward Sgr B2(N1E) with ALMA.

| Rest Freq. (MHz) | Transition $J'_{Ka',Kc'} \leftarrow J''_{Ka'',Kc''}$ : Sym | $E_u$ (K) | $\log_{10} A_{ul}$ (s$^{-1}$) | $S_{ij}\mu^2$ (D$^2$) |
|---|---|---|---|---|
| \multicolumn{5}{c}{Ground $(v_{29}, v_{30}) = (0, 0)$ state} | | | | |
| 87980.402(0.002) | $14_{2,13} - 13_{1,12}$ : $E$ | 33.546 | -4.4914 | 118.0 |
| 87980.454(0.002) | $14_{2,13} - 13_{1,12}$ : $A$ | 33.546 | -4.4914 | 118.0 |
| 92141.928(0.002) | $5_{5,1} - 4_{4,1}$ : $E$ | 12.641 | -4.3465 | 54.4 |
| 92142.666(0.002) | $5_{5,1} - 4_{4,1}$ : $A$ | 12.640 | -4.3457 | 54.5 |
| 92142.895(0.002) | $5_{5,0} - 4_{4,0}$ : $E$ | 12.641 | -4.3465 | 54.4 |
| 92143.080(0.002) | $5_{5,0} - 4_{4,0}$ : $A$ | 12.640 | -4.3457 | 54.5 |
| 92402.509(0.003) | $21_{8,13} - 21_{7,14}$ : $E$ | 94.544 | -4.5244 | 140.0 |
| 92403.069(0.004) | $21_{8,13} - 21_{7,14}$ : $A$ | 94.544 | -4.5243 | 140.0 |
| 92407.406(0.004) | $29_{9,20} - 29_{8,21}$ : $E$ | 166.366 | -4.4713 | 217.0 |
| 92408.136(0.003) | $29_{9,20} - 29_{8,21}$ : $A$ | 166.366 | -4.4713 | 217.0 |
| 102058.096(0.003) | $31_{9,23} - 31_{8,24}$ : $E$ | 186.341 | -4.3451 | 230.0 |
| 102058.653(0.003) | $31_{9,23} - 31_{8,24}$ : $A$ | 186.340 | -4.3451 | 230.0 |
| 103341.999(0.003) | $32_{6,26} - 32_{5,27}$ : $E$ | 186.750 | -4.4031 | 200.0 |
| 103342.235(0.004) | $32_{6,26} - 32_{5,27}$ : $A$ | 186.749 | -4.4031 | 200.0 |

Notes. Col. (1): Rest frequencies with calculated uncertainties given in parentheses; Col. (2): Transition; Col. (3): Upper level energy. For the $v_{30} = 1$ state, the tabulated values take the excitation energy of 45 cm$^{-1}$ into account; Col. (4): Base 10 logarithm of the Einstein $A_{ul}$ coefficient; Col. (5): Line strength $S_{ij}\mu^2$.

Table 4. Parameters of our best-fit LTE model of $C_2H_5CONH_2$, $CH_3CONH_2$, $NH_2CHO$ and $C^{18}O$ toward Sgr B2(N1E) and Sgr B2(N1) with the ALMA telescope.

| Molecule | Size (") | $T_{rot}$ (K) | $N$ ($cm^{-2}$) | $\Delta V$ ($kms^{-1}$) | $V_{off}$ ($kms^{-1}$) |
|---|---|---|---|---|---|
| Sgr B2(N1E) | | | | | |
| $C_2H_5CONH_2$, v=0, v=1 | 2.3 | 150 | 1.5(16) | 4.2(.5)[a] | -1 |
| $CH_3CONH_2$, v=0 | 2.3 | 150 | 2.5(16) | 3.6(.6)[b] | -1 |
| $NH_2CHO$, v=0 | 2.3 | 150 | 2.8(17) | 6[c] | -1 |
| $C^{18}O$ | 2.3 | 150 | 2.5(17) | 3.5(.3) | 1 |
| Sgr B2(N1) | | | | | |
| $C_2H_5CONH_2$, v=0 | 2.3 | 150 | 3.3(16) | 5[d] | -1 |
| $CH_3CONH_2$, v=0 | 2.3 | 150 | 1.5(17) | 5.4(.4)[e] | -1 |
| $NH_2CHO$, v=0 | 2.3 | 150 | 9.5(17) | 7[c] | -1 |

Notes. Col. (1): Molecule name; Col. (2): Source diameter (FWHM), Col. (3): Rotational temperature; Col. (4): Total column density of the molecule. x(y) means $x \times 10^y$; Col. (5): Linewidth (FWHM); Col. (6): Velocity offset with respect to the assumed systemic velocity of Sgr B2(N1), $V_{sys} = 64$ km s$^{-1}$. [a]: Gaussian fitting result of 87.980 and 92.402 GHz transitions. [b]: Gaussian fitting result of 92.338 GHz transitions. [c]: As $NH_2CHO$ is optically thick, the linewidth obtained with gaussian fitting is unreliable, thus the Weeds fitting by visual matching was done. [d]: As $C_2H_5CONH_2$ blends with an U-line toward the whole region of Sgr B2(N1), the linewidth and its uncertainty cannot be derived with GAUSSION fitting in CLASS, thus the Weeds fitting by visual matching was done. [d]: Gaussian fitting result of 92.343 GHz transitions.

Table 5. Parameters of preliminary model of other identified species toward Sgr B2(N1E) with the ALMA telescope.

| Molecule | Size (") | $T_{rot}$ (K) | $N$ ($cm^{-2}$) | $\Delta V$ ($kms^{-1}$) | $V_{off}$ ($kms^{-1}$) |
|---|---|---|---|---|---|
| $CH_3COOH$, vt=0 | 2.3 | 150 | 6.0(17) | 5 | 0 |
| $CH_3COOH$, vt=1 | 2.3 | 150 | 1.0(16) | 5 | 0 |
| i-$C_3H_7CN$ | 2.3 | 150 | 2.0(16) | 5 | -2 |
| n-$C_3H_7CN$, v=0 | 2.3 | 150 | 2.0(16) | 5 | -2 |
| $CH_3O^{13}CHO$, vt=0,1 | 2.3 | 150 | 4.0(17) | 4 | 0 |
| $C_2H_3CN$, v=0 | 2.3 | 150 | 1.0(17) | 7 | -1 |
| $H_2CCHC^{15}N$ | 2.3 | 150 | 1.2(15) | 5 | -1 |
| $H_2CCH^{13}CN$, v=0 | 2.3 | 150 | 1.0(16) | 5 | -1 |
| $H_2^{13}CCHCN$, v=0 | 2.3 | 150 | 4.8(15) | 6 | -1 |
| $C_2H_5OH$, v=0 | 2.3 | 150 | 1.5(17) | 7 | -1 |
| $NH_2CHO$, v=0 | 2.3 | 150 | 2.4(17) | 6 | -1 |
| $C_2H_5CN$, v=0 | 2.3 | 150 | 1.5(17) | 7 | -1 |
| $CH_3^{13}CH_2CN$, v=0 | 2.3 | 150 | 7.0(15) | 5 | -2 |
| $C_2H_5CN$, v20=1-A | 2.3 | 150 | 8.0(16) | 5 | -1 |
| $H_2NCH_2CN$, v=0 | 2.3 | 150 | 5.0(15) | 5 | -1 |
| $CH_2OHCHO$ | 2.3 | 150 | 3.0(15) | 5 | -1 |
| $CH_3CN$, v8=1 | 2.3 | 150 | 8(16) | 6 | 0 |
| $CH_3CN$, v8=2 | 2.3 | 150 | 5(17) | 5 | -1 |
| $CH_3NCO$, vb=0 | 2.3 | 150 | 3.0(16) | 5 | -1 |
| $CH_3OCH_3$, v=0 | 2.3 | 150 | 2.0(16) | 4 | -1 |
| $CH_3SH$, vt=0 | 2.3 | 150 | 2.5(17) | 6 | 0 |
| $SO_2$, v2=1 | 2.3 | 150 | 2.5(18) | 5 | 0 |
| $HOCH_2CN$ | 2.3 | 150 | 5.0(15) | 5 | 0 |
| $CH_3OCHO$ | 2.3 | 150 | 6.0(16) | 4 | 0 |
| $CH_3^{13}CHO$ | 2.3 | 150 | 1.0(18) | 5 | -1 |

Notes. Col. (1): Molecule name; Col. (2): Source diameter (FWHM), Col. (3): Rotational temperature; Col. (4): Total column density of the molecule. x(y) means $x \times 10^y$; Col. (5): Linewidth (FWHM); Col. (6): Velocity offset with respect to the assumed systemic velocity of Sgr B2(N1), $V_{sys} = 64$ km s$^{-1}$.

# APPENDIX

Table A1. List of calculated positions and assignments of rotational transitions in the ground vibrational state (v29,v30) = (0,0) of propionamide up to J = 40 in the 1-460 GHz frequency range.

| (1) Upper level | | | | | | (2) Lower level | | | | | (3) Calculated | (4) Unc. | (5) $E_{low}$ | (6) $S\mu^2$ |
|---|---|---|---|---|---|---|---|---|---|---|---|---|---|---|
| $S'_y$ | $m'$ | $F'$ | $J'$ | $K'_a$ | $K'_c$ | $S_y$ | $m$ | $F$ | $J$ | $K_a$ | $K_c$ | (MHz) | (MHz) | (cm$^{-1}$) | ($D^2$) |
| A2 | 0 | 17.0 | 17 | 6 | 11 | A1 | 0 | 17.0 | 17 | 6 | 12 | 1046.6175 | 0.0004 | 42.1815 | 0.147E+01 |
| E  | 1 | 17.0 | 17 | 6 | 11 | E  | 1 | 17.0 | 17 | 6 | 12 | 1046.6486 | 0.0004 | 42.1817 | 0.147E+01 |
| A2 | 0 | 18.0 | 17 | 6 | 11 | A1 | 0 | 18.0 | 17 | 6 | 12 | 1046.6987 | 0.0004 | 42.1815 | 0.156E+01 |
| A2 | 0 | 16.0 | 17 | 6 | 11 | A1 | 0 | 16.0 | 17 | 6 | 12 | 1046.7035 | 0.0004 | 42.1815 | 0.139E+01 |
| E  | 1 | 18.0 | 17 | 6 | 11 | E  | 1 | 18.0 | 17 | 6 | 12 | 1046.7298 | 0.0004 | 42.1817 | 0.156E+01 |
| E  | 1 | 16.0 | 17 | 6 | 11 | E  | 1 | 16.0 | 17 | 6 | 12 | 1046.7346 | 0.0004 | 42.1817 | 0.139E+01 |
| A1 | 0 | 10.0 | 10 | 4 | 6  | A2 | 0 | 10.0 | 10 | 4 | 7  | 1057.1069 | 0.0004 | 15.7960 | 0.112E+01 |
| E  | 1 | 10.0 | 10 | 4 | 6  | E  | 1 | 10.0 | 10 | 4 | 7  | 1057.1240 | 0.0004 | 15.7963 | 0.112E+01 |
| A1 | 0 | 11.0 | 10 | 4 | 6  | A2 | 0 | 11.0 | 10 | 4 | 7  | 1057.2603 | 0.0004 | 15.7960 | 0.124E+01 |
| A1 | 0 | 9.0  | 10 | 4 | 6  | A2 | 0 | 9.0  | 10 | 4 | 7  | 1057.2758 | 0.0004 | 15.7960 | 0.102E+01 |
| E  | 1 | 11.0 | 10 | 4 | 6  | E  | 1 | 11.0 | 10 | 4 | 7  | 1057.2775 | 0.0004 | 15.7963 | 0.124E+01 |
| E  | 1 | 9.0  | 10 | 4 | 6  | E  | 1 | 9.0  | 10 | 4 | 7  | 1057.2929 | 0.0004 | 15.7963 | 0.102E+01 |
| A1 | 0 | 21.0 | 21 | 7 | 14 | A2 | 0 | 21.0 | 21 | 7 | 15 | 1310.0231 | 0.0005 | 62.5847 | 0.159E+01 |
| E  | 1 | 21.0 | 21 | 7 | 14 | E  | 1 | 21.0 | 21 | 7 | 15 | 1310.0605 | 0.0005 | 62.5849 | 0.159E+01 |
| A1 | 0 | 22.0 | 21 | 7 | 14 | A2 | 0 | 22.0 | 21 | 7 | 15 | 1310.1005 | 0.0005 | 62.5847 | 0.167E+01 |
| A1 | 0 | 20.0 | 21 | 7 | 14 | A2 | 0 | 20.0 | 21 | 7 | 15 | 1310.1042 | 0.0005 | 62.5847 | 0.152E+01 |
| E  | 1 | 22.0 | 21 | 7 | 14 | E  | 1 | 22.0 | 21 | 7 | 15 | 1310.1379 | 0.0005 | 62.5849 | 0.167E+01 |
| E  | 1 | 20.0 | 21 | 7 | 14 | E  | 1 | 20.0 | 21 | 7 | 15 | 1310.1416 | 0.0005 | 62.5849 | 0.152E+01 |
| A2 | 0 | 6.0  | 7  | 3 | 5  | A1 | 0 | 5.0  | 6  | 4 | 2  | 1340.8581 | 0.0021 | 8.1666  | 0.464E+01 |
| E  | 1 | 6.0  | 7  | 3 | 5  | E  | 1 | 5.0  | 6  | 4 | 2  | 1340.9264 | 0.0021 | 8.1669  | 0.458E+01 |
| A2 | 0 | 8.0  | 7  | 3 | 5  | A1 | 0 | 7.0  | 6  | 4 | 2  | 1340.9268 | 0.0021 | 8.1666  | 0.622E+01 |
| E  | 1 | 8.0  | 7  | 3 | 5  | E  | 1 | 7.0  | 6  | 4 | 2  | 1340.9950 | 0.0021 | 8.1669  | 0.613E+01 |
| A2 | 0 | 7.0  | 7  | 3 | 5  | A1 | 0 | 6.0  | 6  | 4 | 2  | 1341.3621 | 0.0021 | 8.1666  | 0.538E+01 |
| E  | 1 | 7.0  | 7  | 3 | 5  | E  | 1 | 6.0  | 6  | 4 | 2  | 1341.4303 | 0.0021 | 8.1669  | 0.530E+01 |
| A1 | 0 | 7.0  | 7  | 3 | 4  | A2 | 0 | 7.0  | 7  | 3 | 5  | 1348.0582 | 0.0004 | 8.2114  | 0.875E+00 |
| E  | 1 | 7.0  | 7  | 3 | 4  | E  | 1 | 7.0  | 7  | 3 | 5  | 1348.0682 | 0.0004 | 8.2116  | 0.875E+00 |
| A1 | 0 | 8.0  | 7  | 3 | 4  | A2 | 0 | 8.0  | 7  | 3 | 5  | 1348.3441 | 0.0004 | 8.2114  | 0.101E+01 |
| E  | 1 | 8.0  | 7  | 3 | 4  | E  | 1 | 8.0  | 7  | 3 | 5  | 1348.3541 | 0.0004 | 8.2116  | 0.101E+01 |
| A1 | 0 | 6.0  | 7  | 3 | 4  | A2 | 0 | 6.0  | 7  | 3 | 5  | 1348.3853 | 0.0004 | 8.2114  | 0.770E+00 |
| E  | 1 | 6.0  | 7  | 3 | 4  | E  | 1 | 6.0  | 7  | 3 | 5  | 1348.3953 | 0.0004 | 8.2116  | 0.770E+00 |
| E  | 1 | 6.0  | 7  | 3 | 5  | E  | 1 | 5.0  | 6  | 4 | 3  | 1358.9304 | 0.0023 | 8.1663  | 0.684E-01 |
| E  | 1 | 8.0  | 7  | 3 | 5  | E  | 1 | 7.0  | 6  | 4 | 3  | 1358.9979 | 0.0023 | 8.1663  | 0.916E-01 |
| E  | 1 | 7.0  | 7  | 3 | 5  | E  | 1 | 6.0  | 6  | 4 | 3  | 1359.4265 | 0.0023 | 8.1663  | 0.792E-01 |
| A1 | 0 | 3.0  | 4  | 1 | 4  | A2 | 0 | 2.0  | 3  | 2 | 1  | 1407.3994 | 0.0014 | 2.2190  | 0.214E+01 |
| E  | 1 | 3.0  | 4  | 1 | 4  | E  | 1 | 2.0  | 3  | 2 | 1  | 1407.5548 | 0.0014 | 2.2193  | 0.214E+01 |
| A1 | 0 | 5.0  | 4  | 1 | 4  | A2 | 0 | 4.0  | 3  | 2 | 1  | 1407.9512 | 0.0013 | 2.2190  | 0.367E+01 |
| E  | 1 | 5.0  | 4  | 1 | 4  | E  | 1 | 4.0  | 3  | 2 | 1  | 1408.1065 | 0.0013 | 2.2193  | 0.367E+01 |
| A1 | 0 | 4.0  | 4  | 1 | 4  | A2 | 0 | 3.0  | 3  | 2 | 1  | 1409.9772 | 0.0016 | 2.2190  | 0.281E+01 |
| E  | 1 | 4.0  | 4  | 1 | 4  | E  | 1 | 3.0  | 3  | 2 | 1  | 1410.1326 | 0.0016 | 2.2193  | 0.281E+01 |
| A1 | 0 | 4.0  | 4  | 2 | 2  | A2 | 0 | 4.0  | 4  | 2 | 3  | 1463.5393 | 0.0005 | 3.0850  | 0.623E+00 |
| E  | 1 | 4.0  | 4  | 2 | 2  | E  | 1 | 4.0  | 4  | 2 | 3  | 1463.5418 | 0.0005 | 3.0853  | 0.623E+00 |
| A1 | 0 | 5.0  | 4  | 2 | 2  | A2 | 0 | 5.0  | 4  | 2 | 3  | 1464.1122 | 0.0003 | 3.0850  | 0.810E+00 |
| E  | 1 | 5.0  | 4  | 2 | 2  | E  | 1 | 5.0  | 4  | 2 | 3  | 1464.1147 | 0.0003 | 3.0853  | 0.810E+00 |
| A1 | 0 | 3.0  | 4  | 2 | 2  | A2 | 0 | 3.0  | 4  | 2 | 3  | 1464.2595 | 0.0004 | 3.0850  | 0.503E+00 |
| E  | 1 | 3.0  | 4  | 2 | 2  | E  | 1 | 3.0  | 4  | 2 | 3  | 1464.2620 | 0.0004 | 3.0853  | 0.503E+00 |
| E  | 1 | 34.0 | 34 | 8 | 27 | E  | 1 | 35.0 | 35 | 5 | 30 | 1522.9726 | 0.0091 | 148.6103 | 0.519E+01 |
| A2 | 0 | 34.0 | 34 | 8 | 27 | A1 | 0 | 35.0 | 35 | 5 | 30 | 1523.6387 | 0.0091 | 148.6101 | 0.519E+01 |
| E  | 1 | 35.0 | 34 | 8 | 27 | E  | 1 | 36.0 | 35 | 5 | 30 | 1523.6874 | 0.0091 | 148.6103 | 0.534E+01 |
| E  | 1 | 33.0 | 34 | 8 | 27 | E  | 1 | 34.0 | 35 | 5 | 30 | 1523.7077 | 0.0091 | 148.6103 | 0.504E+01 |
| A2 | 0 | 35.0 | 34 | 8 | 27 | A1 | 0 | 36.0 | 35 | 5 | 30 | 1524.3535 | 0.0091 | 148.6101 | 0.534E+01 |

Notes. Col. (1): Quantum numbers for upper energy levels; Col. (2): Quantum numbers for lower energy levels, Col. (3): Calculated frequencies; Col. (4): Uncertainties for frequencies; Col. (5): Lower energy levels; Col. (6): Line strength. Full table is available as electronic supplementary material with this article.

# Table A2. List of calculated positions and assignments of rotational transitions in the first excited skeletal torsion state (v29,v30) = (0,1) of propionamide up to J = 30 in the 1-150 GHz frequency range.

| (1) Upper level | | | | | | (2) Lower level | | | | | (3) Calculated | (4) Unc. | (5) $E_{low}$ | (6) $S\mu^2$ |
|---|---|---|---|---|---|---|---|---|---|---|---|---|---|---|
| $S'_y$ | $m'$ | $F'$ | $J'$ | $K'_a$ | $K'_c$ | $S_y$ | $m$ | $F$ | $J$ | $K_a$ | $K_c$ | (MHz) | (MHz) | (cm$^{-1}$) | ($D^2$) |
| A2 | 0 | 17.0 | 17 | 6 | 11 | A1 | 0 | 17.0 | 17 | 6 | 12 | 1029.8309 | 0.0014 | 87.1110 | 0.155E+01 |
| E  | 1 | 17.0 | 17 | 6 | 11 | E  | 1 | 17.0 | 17 | 6 | 12 | 1029.8334 | 0.0013 | 87.1110 | 0.155E+01 |
| A2 | 0 | 18.0 | 17 | 6 | 11 | A1 | 0 | 18.0 | 17 | 6 | 12 | 1029.9102 | 0.0012 | 87.1110 | 0.165E+01 |
| E  | 1 | 18.0 | 17 | 6 | 11 | E  | 1 | 18.0 | 17 | 6 | 12 | 1029.9127 | 0.0012 | 87.1110 | 0.165E+01 |
| A2 | 0 | 16.0 | 17 | 6 | 11 | A1 | 0 | 16.0 | 17 | 6 | 12 | 1029.9149 | 0.0013 | 87.1110 | 0.147E+01 |
| E  | 1 | 16.0 | 17 | 6 | 11 | E  | 1 | 16.0 | 17 | 6 | 12 | 1029.9174 | 0.0012 | 87.1110 | 0.147E+01 |
| A1 | 0 | 10.0 | 10 | 4 | 6  | A2 | 0 | 10.0 | 10 | 4 | 7  | 1039.7338 | 0.0017 | 60.7602 | 0.118E+01 |
| E  | 1 | 10.0 | 10 | 4 | 6  | E  | 1 | 10.0 | 10 | 4 | 7  | 1039.7353 | 0.0017 | 60.7602 | 0.118E+01 |
| A1 | 0 | 11.0 | 10 | 4 | 6  | A2 | 0 | 11.0 | 10 | 4 | 7  | 1039.8837 | 0.0012 | 60.7602 | 0.130E+01 |
| E  | 1 | 11.0 | 10 | 4 | 6  | E  | 1 | 11.0 | 10 | 4 | 7  | 1039.8851 | 0.0011 | 60.7602 | 0.130E+01 |
| A1 | 0 | 9.0  | 10 | 4 | 6  | A2 | 0 | 9.0  | 10 | 4 | 7  | 1039.8987 | 0.0013 | 60.7602 | 0.108E+01 |
| E  | 1 | 9.0  | 10 | 4 | 6  | E  | 1 | 9.0  | 10 | 4 | 7  | 1039.9002 | 0.0012 | 60.7602 | 0.108E+01 |
| E  | 1 | 9.0  | 9  | 2 | 7  | E  | 1 | 10.0 | 10 | 1 | 10 | 1045.0356 | 0.0254 | 56.3124 | 0.978E+00 |
| A2 | 0 | 9.0  | 9  | 2 | 7  | A1 | 0 | 10.0 | 10 | 1 | 10 | 1045.0603 | 0.0258 | 56.3124 | 0.978E+00 |
| E  | 1 | 10.0 | 9  | 2 | 7  | E  | 1 | 11.0 | 10 | 1 | 10 | 1047.5220 | 0.0155 | 56.3124 | 0.108E+01 |
| A2 | 0 | 10.0 | 9  | 2 | 7  | A1 | 0 | 11.0 | 10 | 1 | 10 | 1047.5467 | 0.0161 | 56.3124 | 0.108E+01 |
| E  | 1 | 8.0  | 9  | 2 | 7  | E  | 1 | 9.0  | 10 | 1 | 10 | 1047.7753 | 0.0175 | 56.3124 | 0.884E+00 |
| A2 | 0 | 8.0  | 9  | 2 | 7  | A1 | 0 | 9.0  | 10 | 1 | 10 | 1047.8000 | 0.0181 | 56.3124 | 0.884E+00 |
| A1 | 0 | 21.0 | 21 | 7 | 14 | A2 | 0 | 21.0 | 21 | 7 | 15 | 1289.4397 | 0.0016 | 107.4964 | 0.168E+01 |
| A1 | 0 | 21.0 | 21 | 7 | 14 | A2 | 0 | 21.0 | 21 | 7 | 15 | 1289.4397 | 0.0016 | 107.4964 | 0.168E+01 |
| A1 | 0 | 21.0 | 21 | 7 | 14 | A2 | 0 | 21.0 | 21 | 7 | 15 | 1289.4397 | 0.0016 | 107.4964 | 0.168E+01 |
| E  | 1 | 21.0 | 21 | 7 | 14 | E  | 1 | 21.0 | 21 | 7 | 15 | 1289.4433 | 0.0015 | 107.4964 | 0.168E+01 |
| A1 | 0 | 22.0 | 21 | 7 | 14 | A2 | 0 | 22.0 | 21 | 7 | 15 | 1289.5153 | 0.0015 | 107.4964 | 0.176E+01 |
| E  | 1 | 22.0 | 21 | 7 | 14 | E  | 1 | 22.0 | 21 | 7 | 15 | 1289.5189 | 0.0014 | 107.4964 | 0.176E+01 |
| A1 | 0 | 20.0 | 21 | 7 | 14 | A2 | 0 | 20.0 | 21 | 7 | 15 | 1289.5190 | 0.0015 | 107.4964 | 0.160E+01 |
| E  | 1 | 20.0 | 21 | 7 | 14 | E  | 1 | 20.0 | 21 | 7 | 15 | 1289.5225 | 0.0014 | 107.4964 | 0.160E+01 |
| A1 | 0 | 7.0  | 7  | 3 | 4  | A2 | 0 | 7.0  | 7  | 3 | 5  | 1325.7526 | 0.0028 | 53.1903 | 0.922E+00 |
| E  | 1 | 7.0  | 7  | 3 | 4  | E  | 1 | 7.0  | 7  | 3 | 5  | 1325.7535 | 0.0028 | 53.1903 | 0.922E+00 |
| A1 | 0 | 8.0  | 7  | 3 | 4  | A2 | 0 | 8.0  | 7  | 3 | 5  | 1326.0316 | 0.0014 | 53.1903 | 0.107E+01 |
| E  | 1 | 8.0  | 7  | 3 | 4  | E  | 1 | 8.0  | 7  | 3 | 5  | 1326.0325 | 0.0014 | 53.1903 | 0.107E+01 |
| A1 | 0 | 6.0  | 7  | 3 | 4  | A2 | 0 | 6.0  | 7  | 3 | 5  | 1326.0718 | 0.0019 | 53.1903 | 0.812E+00 |
| E  | 1 | 6.0  | 7  | 3 | 4  | E  | 1 | 6.0  | 7  | 3 | 5  | 1326.0728 | 0.0019 | 53.1903 | 0.812E+00 |
| E  | 1 | 13.0 | 13 | 8 | 6  | E  | 1 | 14.0 | 14 | 7 | 7  | 1380.8726 | 0.0271 | 78.9832 | 0.387E+01 |
| E  | 1 | 14.0 | 13 | 8 | 6  | E  | 1 | 15.0 | 14 | 7 | 7  | 1381.0501 | 0.0239 | 78.9832 | 0.416E+01 |
| E  | 1 | 12.0 | 13 | 8 | 6  | E  | 1 | 13.0 | 14 | 7 | 7  | 1381.0638 | 0.0243 | 78.9832 | 0.360E+01 |
| A1 | 0 | 13.0 | 13 | 8 | 6  | A2 | 0 | 14.0 | 14 | 7 | 7  | 1381.6452 | 0.0185 | 78.9831 | 0.970E+01 |
| A1 | 0 | 14.0 | 13 | 8 | 6  | A2 | 0 | 15.0 | 14 | 7 | 7  | 1381.8227 | 0.0131 | 78.9831 | 0.104E+02 |
| A1 | 0 | 12.0 | 13 | 8 | 6  | A2 | 0 | 13.0 | 14 | 7 | 7  | 1381.8364 | 0.0139 | 78.9831 | 0.903E+01 |
| E  | 1 | 13.0 | 13 | 8 | 5  | E  | 1 | 14.0 | 14 | 7 | 7  | 1382.1497 | 0.0211 | 78.9832 | 0.583E+01 |
| E  | 1 | 14.0 | 13 | 8 | 5  | E  | 1 | 15.0 | 14 | 7 | 7  | 1382.3272 | 0.0165 | 78.9832 | 0.626E+01 |
| E  | 1 | 12.0 | 13 | 8 | 5  | E  | 1 | 13.0 | 14 | 7 | 7  | 1382.3409 | 0.0172 | 78.9832 | 0.543E+01 |
| E  | 1 | 13.0 | 13 | 8 | 6  | E  | 1 | 14.0 | 14 | 7 | 8  | 1385.7711 | 0.0228 | 78.9830 | 0.583E+01 |
| E  | 1 | 14.0 | 13 | 8 | 6  | E  | 1 | 15.0 | 14 | 7 | 8  | 1385.9493 | 0.0188 | 78.9830 | 0.626E+01 |
| E  | 1 | 12.0 | 13 | 8 | 6  | E  | 1 | 13.0 | 14 | 7 | 8  | 1385.9630 | 0.0193 | 78.9830 | 0.543E+01 |
| A2 | 0 | 13.0 | 13 | 8 | 5  | A1 | 0 | 14.0 | 14 | 7 | 8  | 1386.4487 | 0.0185 | 78.9830 | 0.970E+01 |
| A2 | 0 | 14.0 | 13 | 8 | 5  | A1 | 0 | 15.0 | 14 | 7 | 8  | 1386.6269 | 0.0131 | 78.9830 | 0.104E+02 |
| A2 | 0 | 12.0 | 13 | 8 | 5  | A1 | 0 | 13.0 | 14 | 7 | 8  | 1386.6406 | 0.0139 | 78.9830 | 0.903E+01 |
| E  | 1 | 13.0 | 13 | 8 | 5  | E  | 1 | 14.0 | 14 | 7 | 8  | 1387.0482 | 0.0251 | 78.9830 | 0.387E+01 |
| E  | 1 | 14.0 | 13 | 8 | 5  | E  | 1 | 15.0 | 14 | 7 | 8  | 1387.2264 | 0.0213 | 78.9830 | 0.416E+01 |
| E  | 1 | 12.0 | 13 | 8 | 5  | E  | 1 | 13.0 | 14 | 7 | 8  | 1387.2401 | 0.0218 | 78.9830 | 0.360E+01 |
| A1 | 0 | 4.0  | 4  | 2 | 2  | A2 | 0 | 4.0  | 4  | 2 | 3  | 1439.2088 | 0.0054 | 48.0749 | 0.656E+00 |
| E  | 1 | 4.0  | 4  | 2 | 2  | E  | 1 | 4.0  | 4  | 2 | 3  | 1439.2090 | 0.0054 | 48.0750 | 0.656E+00 |

Notes. Col. (1): Quantum numbers for upper energy levels; Col. (2): Quantum numbers for lower energy levels, Col. (3): Calculated frequencies; Col. (4): Uncertainties for frequencies; Col. (5): Lower energy levels; Col. (6): Line strength. Full table is available as electronic supplementary material with this article.

Table A3. Assignments, measured transition frequencies, and residuals from the fit of the microwave, millimeter-wave, and submillimeter-wave (v29,v30) = (0,0) and (1,0) data for propionamide.

| (1) Upper level | | | | | | (2) Lower level | | | | | (3) Measured | (4) Unc. | (5) obs.-calc | (6) Source of data |
|---|---|---|---|---|---|---|---|---|---|---|---|---|---|---|
| $S'_y$ | $m'$ | $F'$ | $J'$ | $K'_a$ | $K'_c$ | $S_y$ | $m$ | $F$ | $J$ | $K_a$ | $K_c$ | (MHz) | (MHz) | (MHz) | |
| E  | 1 | 3.0 | 3 | 1 | 2 | E  | 1 | 3.0 | 3 | 0 | 3 | 9828.3293  | 0.0040 | -0.0012 | NIST |
| A1 | 0 | 3.0 | 3 | 1 | 2 | A2 | 0 | 3.0 | 3 | 0 | 3 | 9828.3967  | 0.0040 | 0.0052  | NIST |
| E  | 1 | 4.0 | 3 | 1 | 2 | E  | 1 | 3.0 | 3 | 0 | 3 | 9828.7609  | 0.0040 | -0.0040 | NIST |
| A1 | 0 | 4.0 | 3 | 1 | 2 | A2 | 0 | 3.0 | 3 | 0 | 3 | 9828.8307  | 0.0040 | 0.0048  | NIST |
| A1 | 0 | 2.0 | 3 | 1 | 2 | A2 | 0 | 3.0 | 3 | 0 | 3 | 9828.9850  | 0.0080 | 0.0071  | NIST |
| E  | 1 | 4.0 | 3 | 1 | 2 | E  | 1 | 4.0 | 3 | 0 | 3 | 9829.7715  | 0.0040 | -0.0007 | NIST |
| A1 | 0 | 4.0 | 3 | 1 | 2 | A2 | 0 | 4.0 | 3 | 0 | 3 | 9829.8431  | 0.0040 | 0.0099  | NIST |
| E  | 1 | 2.0 | 3 | 1 | 2 | E  | 1 | 2.0 | 3 | 0 | 3 | 9830.2738  | 0.0040 | -0.0030 | NIST |
| A1 | 0 | 2.0 | 3 | 1 | 2 | A2 | 0 | 2.0 | 3 | 0 | 3 | 9830.3444  | 0.0040 | 0.0066  | NIST |
| A2 | 0 | 4.0 | 4 | 1 | 3 | A1 | 0 | 3.0 | 3 | 2 | 2 | 11341.7526 | 0.0040 | 0.0004  | NIST |
| A2 | 0 | 4.0 | 4 | 1 | 3 | A1 | 0 | 4.0 | 3 | 2 | 2 | 11341.7526 | 0.0040 | 0.0004  | NIST |
| E  | 1 | 4.0 | 4 | 1 | 3 | E  | 1 | 3.0 | 3 | 2 | 2 | 11341.8796 | 0.0040 | -0.0005 | NIST |
| E  | 1 | 4.0 | 4 | 1 | 3 | E  | 1 | 4.0 | 3 | 2 | 2 | 11341.8796 | 0.0040 | -0.0005 | NIST |
| A2 | 0 | 5.0 | 4 | 1 | 3 | A1 | 0 | 4.0 | 3 | 2 | 2 | 11342.0490 | 0.0040 | 0.0002  | NIST |
| E  | 1 | 5.0 | 4 | 1 | 3 | E  | 1 | 4.0 | 3 | 2 | 2 | 11342.1753 | 0.0040 | -0.0015 | NIST |
| A1 | 0 | 1.0 | 2 | 1 | 2 | A2 | 0 | 1.0 | 1 | 1 | 1 | 12324.3051 | 0.0040 | 0.0000  | NIST |
| E  | 1 | 1.0 | 2 | 1 | 2 | E  | 1 | 1.0 | 1 | 1 | 1 | 12324.3051 | 0.0040 | 0.0026  | NIST |
| A1 | 0 | 1.0 | 2 | 1 | 2 | A2 | 0 | 2.0 | 1 | 1 | 1 | 12324.8907 | 0.0040 | 0.0006  | NIST |
| E  | 1 | 1.0 | 2 | 1 | 2 | E  | 1 | 2.0 | 1 | 1 | 1 | 12324.8907 | 0.0040 | 0.0032  | NIST |
| A1 | 0 | 3.0 | 2 | 1 | 2 | A2 | 0 | 2.0 | 1 | 1 | 1 | 12325.5987 | 0.0040 | -0.0015 | NIST |
| E  | 1 | 3.0 | 2 | 1 | 2 | E  | 1 | 2.0 | 1 | 1 | 1 | 12325.5987 | 0.0040 | 0.0011  | NIST |
| A1 | 0 | 1.0 | 2 | 1 | 2 | A2 | 0 | 0.0 | 1 | 1 | 1 | 12325.7607 | 0.0040 | -0.0068 | NIST |
| E  | 1 | 1.0 | 2 | 1 | 2 | E  | 1 | 0.0 | 1 | 1 | 1 | 12325.7607 | 0.0040 | -0.0041 | NIST |
| A1 | 0 | 2.0 | 2 | 1 | 2 | A2 | 0 | 1.0 | 1 | 1 | 1 | 12326.2924 | 0.0040 | -0.0012 | NIST |
| E  | 1 | 2.0 | 2 | 1 | 2 | E  | 1 | 1.0 | 1 | 1 | 1 | 12326.2924 | 0.0040 | 0.0014  | NIST |

Notes. Col. (1): Quantum numbers for upper energy levels; Col. (2): Quantum numbers for lower energy levels, Col. (3): Measured frequencies; Col. (4): Measurement uncertainties for frequencies; Col. (5): Residuals from the fit; Col. (6): Source of the data. Full table is available as electronic supplementary material with this article.

Table A4. Assignments, measured transition frequencies, and residuals from the fit of the microwave, millimeter-wave, and submillimeter-wave (v29,v30) = (0,1) and (1,1) data for propionamide.

| (1) Upper level | | | | | | (2) Lower level | | | | | (3) Measured | (4) Unc. | (5) obs.-calc | (6) Source of data |
|---|---|---|---|---|---|---|---|---|---|---|---|---|---|---|
| $S'_y$ | $m'$ | $F'$ | $J'$ | $K'_a$ | $K'_c$ | $S_y$ | $m$ | $F$ | $J$ | $K_a$ | $K_c$ | (MHz) | (MHz) | (MHz) | |
| A1 | 0 | -1.0 | 5 | 2 | 4 | A2 | 0 | -1.0 | 5 | 1 | 5 | 26623.9100 | 0.2000 | -0.0432 | OSLO96 |
| E | 1 | -1.0 | 5 | 2 | 4 | E | 1 | -1.0 | 5 | 1 | 5 | 26623.9100 | 0.2000 | -0.0209 | OSLO96 |
| A1 | 0 | -1.0 | 7 | 3 | 4 | A2 | 0 | -1.0 | 7 | 2 | 5 | 26904.5800 | 0.2000 | 0.1708 | OSLO96 |
| E | 1 | -1.0 | 7 | 3 | 4 | E | 1 | -1.0 | 7 | 2 | 5 | 26904.5800 | 0.2000 | 0.2008 | OSLO96 |
| A2 | 0 | -1.0 | 12 | 3 | 9 | A1 | 0 | -1.0 | 12 | 2 | 10 | 27691.6800 | 0.2000 | -0.2383 | OSLO96 |
| E | 1 | -1.0 | 12 | 3 | 9 | E | 1 | -1.0 | 12 | 2 | 10 | 27691.6800 | 0.2000 | -0.2216 | OSLO96 |
| A2 | 0 | -1.0 | 6 | 3 | 3 | A1 | 0 | -1.0 | 6 | 2 | 4 | 28557.9000 | 0.2000 | 0.1453 | OSLO96 |
| E | 1 | -1.0 | 6 | 3 | 3 | E | 1 | -1.0 | 6 | 2 | 4 | 28557.9000 | 0.2000 | 0.1764 | OSLO96 |
| A2 | 0 | -1.0 | 6 | 1 | 5 | A1 | 0 | -1.0 | 5 | 2 | 4 | 28863.8800 | 0.2000 | 0.0671 | OSLO96 |
| E | 1 | -1.0 | 6 | 1 | 5 | E | 1 | -1.0 | 5 | 2 | 4 | 28863.8800 | 0.2000 | 0.0586 | OSLO96 |
| A1 | 0 | -1.0 | 5 | 3 | 2 | A2 | 0 | -1.0 | 5 | 2 | 3 | 30039.3300 | 0.2000 | 0.1117 | OSLO96 |
| E | 1 | -1.0 | 5 | 3 | 2 | E | 1 | -1.0 | 5 | 2 | 3 | 30039.3300 | 0.2000 | 0.1431 | OSLO96 |
| A2 | 0 | -1.0 | 8 | 1 | 7 | A1 | 0 | -1.0 | 8 | 0 | 8 | 31232.8600 | 0.2000 | -0.0066 | OSLO96 |
| E | 1 | -1.0 | 8 | 1 | 7 | E | 1 | -1.0 | 8 | 0 | 8 | 31232.8600 | 0.2000 | 0.0101 | OSLO96 |
| A1 | 0 | -1.0 | 13 | 3 | 10 | A2 | 0 | -1.0 | 13 | 2 | 11 | 31288.2100 | 0.2000 | -0.0560 | OSLO96 |
| E | 1 | -1.0 | 13 | 3 | 10 | E | 1 | -1.0 | 13 | 2 | 11 | 31288.2100 | 0.2000 | -0.0414 | OSLO96 |
| A1 | 0 | -1.0 | 14 | 4 | 10 | A2 | 0 | -1.0 | 14 | 3 | 11 | 32071.8600 | 0.2000 | 0.0986 | OSLO96 |
| E | 1 | -1.0 | 14 | 4 | 10 | E | 1 | -1.0 | 14 | 3 | 11 | 32071.8600 | 0.2000 | 0.1290 | OSLO96 |
| A2 | 0 | -1.0 | 13 | 4 | 9 | A1 | 0 | -1.0 | 13 | 3 | 10 | 32660.0100 | 0.2000 | 0.1812 | OSLO96 |
| E | 1 | -1.0 | 13 | 4 | 9 | E | 1 | -1.0 | 13 | 3 | 10 | 32660.0100 | 0.2000 | 0.2165 | OSLO96 |
| A2 | 0 | -1.0 | 15 | 4 | 11 | A1 | 0 | -1.0 | 15 | 3 | 12 | 32723.9900 | 0.2000 | -0.0986 | OSLO96 |
| E | 1 | -1.0 | 15 | 4 | 11 | E | 1 | -1.0 | 15 | 3 | 12 | 32723.9900 | 0.2000 | -0.0733 | OSLO96 |
| A2 | 0 | -1.0 | 5 | 3 | 3 | A1 | 0 | -1.0 | 5 | 2 | 4 | 33023.3400 | 0.2000 | -0.0966 | OSLO96 |
| E | 1 | -1.0 | 5 | 3 | 3 | E | 1 | -1.0 | 5 | 2 | 4 | 33023.3400 | 0.2000 | -0.0644 | OSLO96 |
| A1 | 0 | -1.0 | 7 | 2 | 6 | A2 | 0 | -1.0 | 7 | 1 | 7 | 33056.5200 | 0.2000 | -0.1523 | OSLO96 |

Notes. Col. (1): Quantum numbers for upper energy levels; Col. (2): Quantum numbers for lower energy uncertainties for frequencies; Col. (5): Residuals from the fit; Col. (6): Source of the data. Full table is available as electronic supplementary material with this article.